\pdfoutput=1
\documentclass[11pt]{article}
\usepackage{extsizes}
\usepackage[letterpaper, total={6in, 8in}]{geometry}
\usepackage{amsmath, amsthm, amssymb, graphicx, dcolumn, bm, array, cancel, esint, color}
\usepackage{epstopdf}
\usepackage{hyperref}
\usepackage{lipsum}
\usepackage{widetext}
\usepackage{tikz, 3dplot}
\usetikzlibrary{backgrounds}
\usepackage{courier}
\usepackage{afterpage}
\usepackage{booktabs}
\usepackage{authblk}
\usepackage[toc,page]{appendix}
\usepackage[sort, numbers]{natbib}
\usepackage{scrextend}
\deffootnote[1.8em]{0pt}{1.6em}{\makebox[1.8em][l]{\thefootnotemark.}}
\hypersetup{
     colorlinks   	= true,
     citecolor   	= red,
     linkcolor       = blue,
     linktoc		= all
}
\definecolor{pink}{RGB}{255, 94, 170}
\begin{document}
\title{\textbf{{Enhancing Low--Cost Ozone Spectrometers\\ to Measure Mesospheric Winds and Tides}}}

\author{O. B. Alam\footnote{Affiliation: \textit{School of Applied and Engineering Physics}; email address: \href{mailto:oba3@cornell.edu}{oba3@cornell.edu}}\\ \textit{\small{Cornell University, Ithaca, New York 14853}}
\and
A. E. E. Rogers\footnote{Corresponding author email address: \href{aeer@haystack.mit.edu}{arogers@haystack.mit.edu}}\\ \textit{\small{MIT Haystack Observatory, Westford, Massachusetts 01886}}}

\date{August 16, 2015}
\maketitle
\abstract{
Ground--based spectrometers have been developed to measure the concentration, velocity, and temperature of ozone in the mesosphere and lower thermosphere (MLT) using low--cost satellite television electronics to observe the 11.072 GHz spectral line of ozone. A two--channel spectrometer has been engineered to yield various performance improvements, including a doubling of the signal--to--noise ratio, improved data processing efficiency, and lower power consumption at 15 W. Following 2009 and 2012 observations of the seasonal and diurnal variations in ozone concentration near the mesopause, the ozone line was observed at an altitude near 95 km and latitude of 38 degrees north using three single--channel spectrometers located at the MIT Haystack Observatory (Westford, MA), Chelmsford High School (Chelmsford, MA), and Union College (Schenectady, NY) pointed south at 8 degrees. Observations from 2009 through 2014 are used to derive the nightly--averaged seasonal variation in meridional velocity, as well as the seasonally--averaged variation with local solar time. The results indicate a seasonal trend in which the winds at 95 km come from the north at about $10~\text{m}\text{s}^{-1}$ in the summer of the northern hemisphere, and from the south at about $10~\text{m}\text{s}^{-1}$ in the winter. Nighttime data from --5 to +5 hours local solar time show a gradual transition of the meridional wind velocity from about --$20~\text{m}\text{s}^{-1}$ to +$20~\text{m}\text{s}^{-1}$. These two trends correlate with nighttime wind measurements from the Millstone Hill High--Resolution F\'abry--Perot Interferometer (FPI) in Westford, MA, which uses the 557.7 nm green line nightglow from atomic oxygen centered at 95 km. The results have also been compared with average meridional winds measured with meteor radar.}
\newpage
\tableofcontents
\newpage
\listoftables
\listoffigures
\newpage
\section{Introduction}
\subsection{Past and New Work}
The \textbf{M}esospheric \textbf{O}zone \textbf{S}ystem for \textbf{A}tmospheric \textbf{I}nvestigations in the \textbf{C}lassroom (MOSAIC) is an array of single--channel radio telescopes that observe the $11.072~4545$ GHz spectral line of ozone in the middle atmosphere. System hardware is based on the Very Small Radio Telescope (VSRT), which was developed at the Massachusetts Institute of Technology (MIT) Haystack Observatory to perform radio observations of the Sun and demonstrate physical principles related to interferometry and electromagnetic radiation. Although the VSRT and single--channel MOSAIC systems were originally intended to be educational outreach tools, they have since been deployed to conduct original scientific research measuring the diurnal, seasonal, and latitudinal variations in ozone concentration \cite{Rogers2009}\cite{Rogers2012}. In recent years, it was realized that a greater resolution of the spectral line was needed in order to produce more reliable data about the ozone's velocity and temperature. A six--channel unit was engineered in order to improve the signal--to--noise ratio by $\sqrt{6}$, and has been in operation for less than one year \cite{True2014}. This technical report encompasses the engineering of an improved two--channel system \cite{VSRT77}. It also presents a detailed study of ozone velocity and preliminary observations of ozone temperature as obtained from a network of single--channel spectrometers observing the spectral line at 95 km. Additionally, we compare our results with the 557.7 nm oxygen spectral line, as well as average winds measured with meteor radar and various theoretical and empirical models of the middle atmosphere.

\subsection{Ozone Dynamics in the Middle Atmosphere}
\begin{figure}[t]
		\centering
		\includegraphics[scale=0.5]{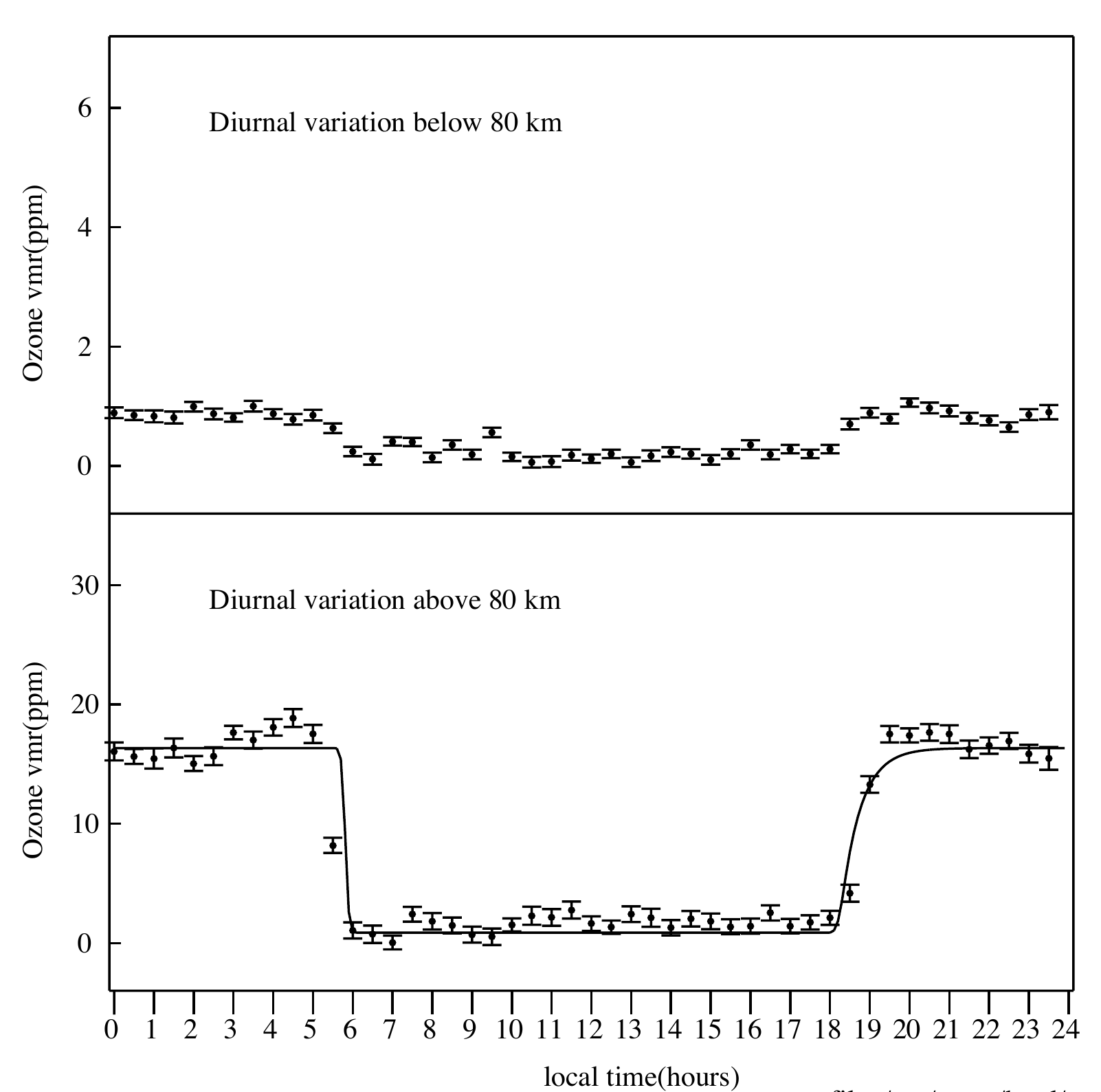}
		\caption{Daytime ultraviolet annihilation of ozone.}
		\label{ozdestroy}
	\end{figure}

The ozone molecule consists of three oxygen atoms ($\text{O}_3$) and is a gas that thrives in various regions of the Earth's atmosphere. The vast majority of ozone resides in the stratosphere, which is 10--50 km from the Earth's surface, and forms the ``ozone layer'' that shields the Earth from solar ultraviolet radiation \cite{OzoneBook}. In much smaller quantities, ozone exists in the mesosphere, which is 50--80 km from the surface, as well as the mesopause, the region between 80-85 km and 100--105 km \cite{Gerding2008} that acts a minimum temperature boundary between the mesosphere and the thermosphere. Although the thermosphere itself is the region above 85 km \cite{IUPAC2014}, we shall refer to our target altitude of 95 km as being ``near the mesopause." As a whole, the region of interest is denoted the ``mesosphere and lower thermosphere'' (MLT) \cite{Davis2013}\cite{Sandford2010}. In this region, new ozone molecules are created through only one atmospheric chemical process:
	\begin{equation}
		\text{O} + \text{O}_2 + \text{M} \rightarrow \text{O}_3 + \text{M},
	\end{equation}
which involves a collision between atomic oxygen ($\text{O}$), diatomic oxygen ($\text{O}_2$), and a third element $\text{M}$, which usually represents diatomic nitrogen ($\text{N}_2$) or diatomic oxygen. There are two processes that destroy ozone in the mesopause, the first being a collision with a single photon that has energy $E=\hbar\omega,$ where $\hbar = 1.055\times 10^{-34}\text{ J}\cdot\text{s}$ is the reduced Planck constant, and $\omega$ is angular frequency in the ultraviolet band, $750\text{ THz}<\omega/2\pi<3000\text{ THz}$. This event splits ozone into atomic and diatomic oxygen:
	\begin{equation}
		\text{O}_3 + \hbar\omega \rightarrow \text{O} + \text{O}_2.
	\end{equation}
The second destructive process is a collision with atomic hydrogen, which produces the hydroxyl free radical (\text{OH}) and diatomic oxygen:
	\begin{equation}
		\text{O}_3 + \text{H} \rightarrow \text{OH} + \text{O}_2. 
	\end{equation}
During the daytime, almost all of the ozone in the mesopause is destroyed by ultraviolet photon collisions. This results in a significant decrease in ozone concentration, followed by a nighttime increase \cite{Rogers2009}\cite{Rogers2012}\cite{VSRT40}\cite{VSRT58}. This is shown in \textcolor{blue}{Figure} \ref{ozdestroy}, where the ozone volume mixing ratio (in parts per million) drops to zero between +6 and +18 hours local solar time. The data were taken from the ozone spectrometer located at MIT Haystack Observatory ($42.618^\circ$ N, $71.498^\circ$ W) and averaged for the first 204 days of 2015.
\subsection{Atmospheric Winds and Tides}	
As microwaves propagate at the speed of light, $c,$ the Doppler shift of the spectral line from one frequency $f_0$ to another,
\begin{equation}\label{Dp}
	f  = \left(1+\frac{v}{c}\right)f_0,
\end{equation}
directly produces the ozone velocity, $v,$ which can reveal information about winds, tides, and other forms of wave propagation in the atmosphere. For example, gravity waves are the result of a gravitationally-driven buoyant restoring force on circulating tropospheric air masses. They dissipate in the stratosphere and mesosphere and heavily influence global air flow \cite{Sandford2010}\cite{PhysAtmos}. Planetary Rossby waves are generated from temperature and pressure fields that arise from the Coriolis effect attributed to the Earth's rotation \cite{PhysAtmos}. However, in the MLT, atmospheric tides are most heavily dominant instead of gravity waves and planetary waves. The solar ultraviolet heating of stratospheric and mesospheric ozone makes atmospheric tides a primary factor in the diurnal trend in ozone velocity. Further importance arises from tidal momentum transfer dynamics, which have been known to be responsible for the coupling between the upper and lower atmosphere \cite{Davis2013}\cite{Sandford2010}. Thus, studying the seasonal and temporal trends in atmospheric tides obtained from wind velocity data are critical to understanding the physics of global air circulation in the middle atmosphere.
\section{The Two--Channel Ozone Spectrometer}
\subsection{Electrical Hardware and Software}	
	\begin{figure}[t]
		\centering
		\centerline{\includegraphics[scale=0.8]{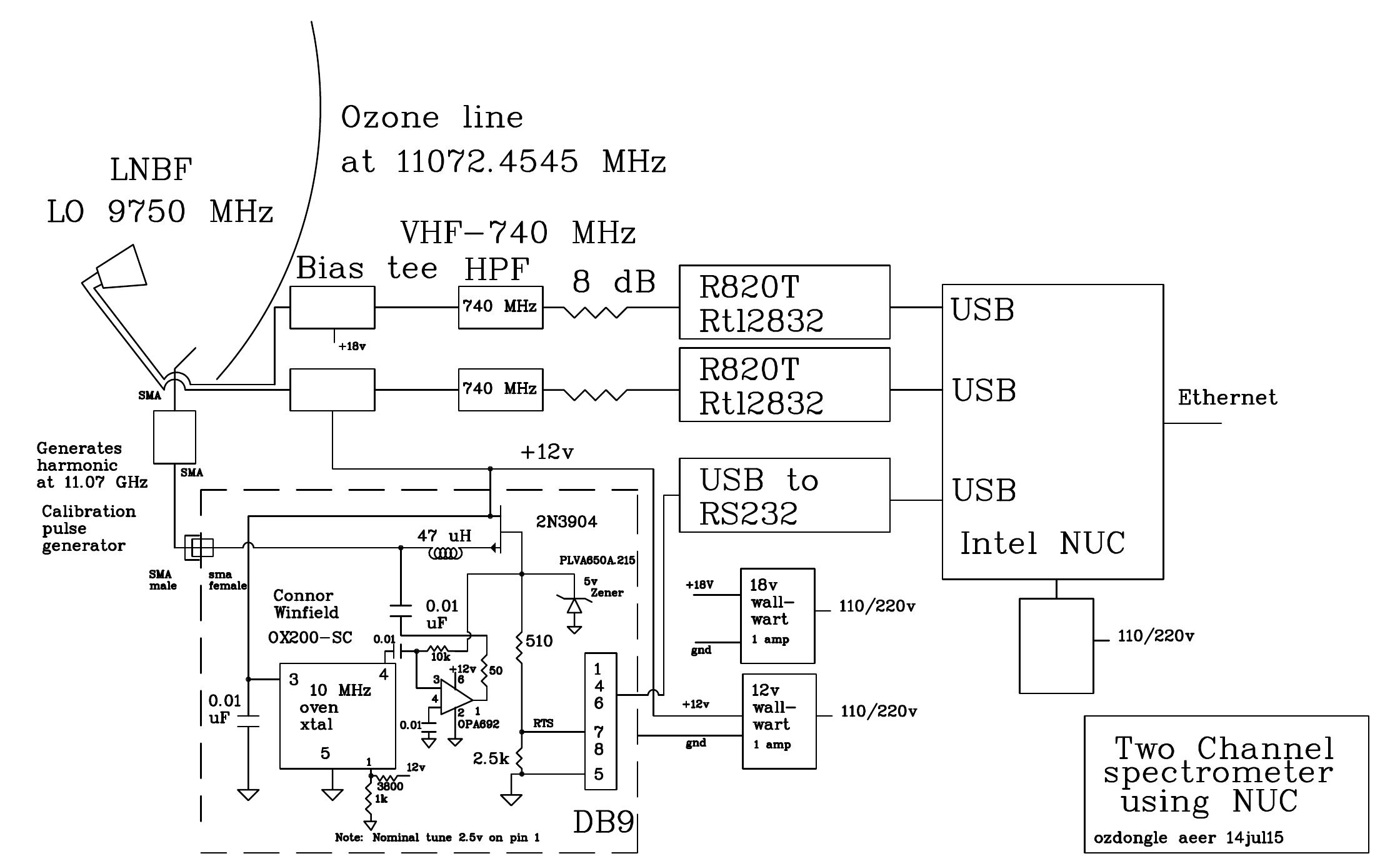}}
		\caption{Two--channel spectrometer block diagram.}
		\label{bdiag}
	\end{figure}		 
	\begin{figure}[t]
		\centering
		\includegraphics[scale=0.4]{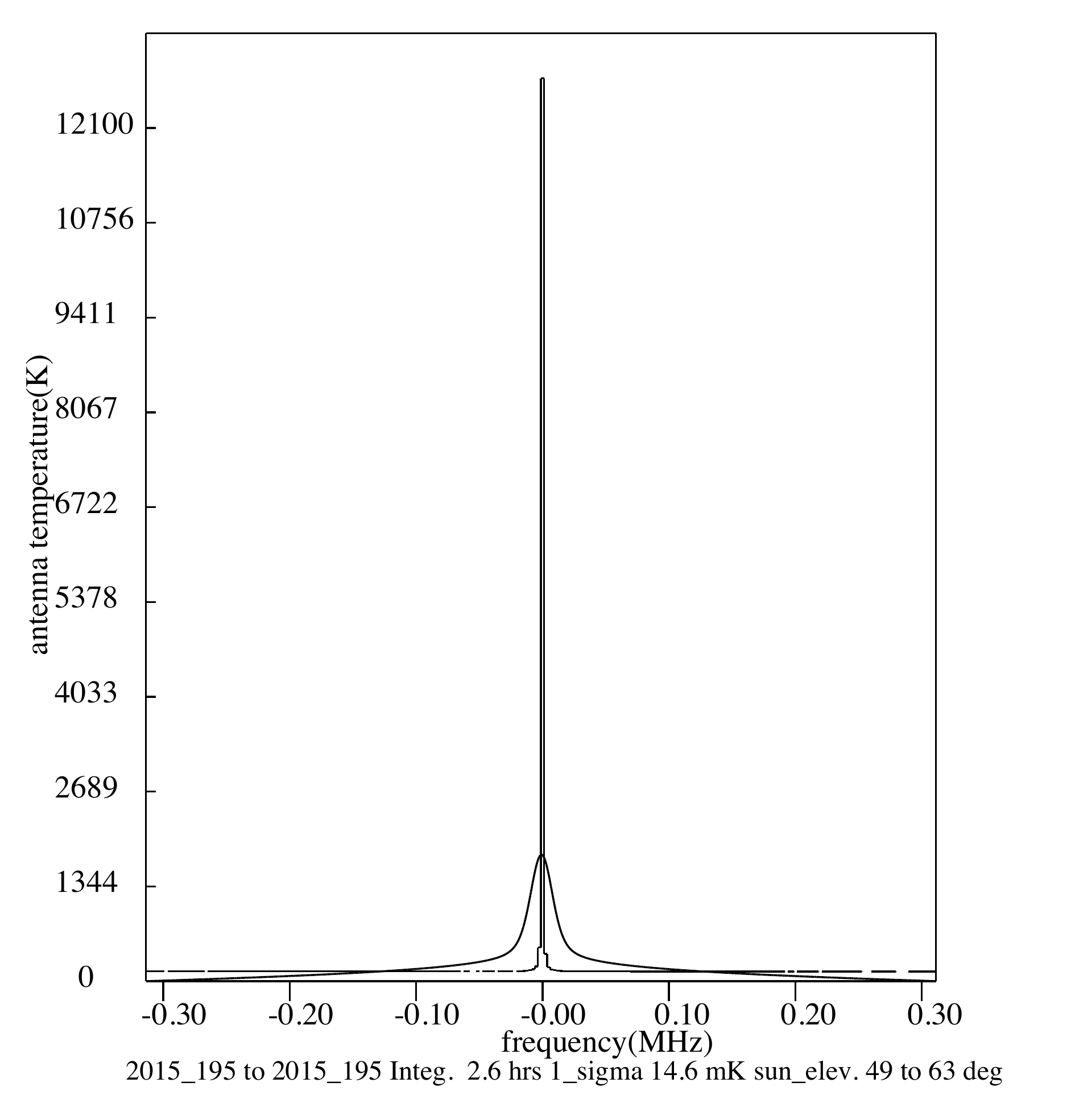}
		\caption{Spectrum of the frequency calibration signal.}
		\label{fcal}
	\end{figure}
The upgraded two--channel MOSAIC hardware (\textcolor{blue}{Figure} \ref{bdiag}) first consists of a Winegard DS-4040 parabolic reflector with 45.72 cm diameter. It is tuned to the 11.072 4545 GHz ozone spectral line by means of a frequency calibration pulse generator, which uses a Connor Winfield OX200-SC 10 MHz oven crystal oscillator to generate a harmonic at 11.070 GHz. It should be noted that the ozone line is in the X band of the microwave spectrum \cite{mbands} but can be resolved with commercially available $\text{K}_\text{u}$ band satellite television electronics, whose waveguides can accept frequencies at the far edge of the X band.\footnote{This is especially true in Europe, where television signals are commonly broadcast at upper X band frequencies around 11 GHz. Thus, an ozone spectrometer in Europe would pick up television interference, while a spectrometer in the United States would be quiet since U.S. television signals are encoded in the $\text{K}_\text{u}$ band above 12 GHz. Because of this distinction, our spectrometers use LNBFs designed for European satellite television dishes.} Specifically, we use a Star--Com SR--3602 \textit{Mini} low-noise block downconverter feed (LNBF) that covers 10.70--11.70 GHz on the low--band and 11.70--12.75 GHz on the high--band \cite{SR3602}. The LNBF has a $9.75$ GHz local oscillator whose signal is heterodyned with the ozone line to downconvert to 1.32 GHz in the L band. The spectrum of the frequency calibration signal (\textcolor{blue}{Figure} \ref{fcal}) is observed by tuning to a line frequency of 1.32 GHz (corresponding to 11.070 GHz) and selecting a command to leave the frequency calibration on \cite{VSRT77}. 

A 12 V wall--wart is used to select for the horizontal microwave polarization (H) on Channel 1, and an 18 V wall--wart is used to select the vertical microwave polarization (V) on Channel 2. Next, in order to minimize out-of-band interference, the signal from each channel is separately fed into a bias tee, followed by a Very High Frequency (VHF) high-pass filter (HPF) at 740 MHz, and finally an 8 dB signal attenuator. Each signal then enters an R820T RTL2832 USB TV dongle that has a local oscillator tuned to the range around 1.42 GHz in order to receive the ozone line. The ozone signal is heterodyned with the local oscillator and then sent to the computer for processing. Our choice of machine is Intel's Next Unit of Computing (NUC), which is small form--factor device, similar to the Raspberry Pi and the BeagleBoard BeagleBone, but more powerful. It runs the Ubuntu Linux 14.04 LTS ``Trusty Tahr" operating system. The source code in C performs a dual--core threaded Fast Fourier Transform (FFT) on the signal to extract the ozone spectrum, and sends the resulting data via Internet to the MOSAIC GUI at \url{www.haystack.mit.edu/ozone/}. For the physical layout of the two--channel spectrometer hardware, see \textcolor{blue}{Figure} \ref{spectr}. 

The optimum digital signal levels after 8 dB attenuation are about mid-scale in the ADC range of -127 to +128. Frequency switching is used, in which the ozone line is alternately placed at 0.9375 and 1.5625 MHz in the baseband\footnote{Equivalently -0.3125 and +0.3125 in the I/Q spectrum.} output. While the line is at 0.9375 MHz, the spectrum at 1.5625 MHz is used as a reference, and vice versa. In addition, the reference spectrum is smoothed using a polynomial. In this case, the noise in the line spectrum is reduced so that the normalized theoretical noise is given by the inverse of the square root of the resolution bandwidth times the integration time as would be obtained by a total power spectrometer. When the spectra of both polarizations are combined, the theoretical noise level for 80 K system noise and 2.44 kHz resolution is 19 mK in one hour. Software timing tests show that for 42 seconds of data ($1.05\times 10^{8}$ complex samples in each channel), the FFT and other calculations add 5 seconds, and the calibration another 10 seconds. Most of the time for calibration is the result of using \texttt{stty} to turn on and off the RS232 RTS line. In principle, this added time could be reduced by using another device, such as a USB power switch dongle.

\subsection{Performance Improvements}
The combination of using both (H) and (V) polarizations and smoothing the reference spectra results in an improvement in the signal--to--noise ratio (SNR) by a factor of 2. In order to further improve the SNR, we selected for the best LNBF by characterizing the noise figures of five different LNBF models. A thorough description of our procedure, and complete results, can be found in \textcolor{blue}{Appendix \S\ref{LNBAP}}. In brief, the Avenger PLS322-S-2 model was determined to have a lower noise figure than the Star--Com SR-3602 \textit{Mini}, but the presence of electronics associated with a phase-locked local oscillator caused spurs in the output spectrum. Our software was unable to correct for rapid spurs, so we chose to continue using the Star--Com SR-3602 \textit{Mini} due to its history of stability and robust downconversion. The typical Y-factor between the beam at 8 degrees elevation and an absorber was 4.9 dB when using the Star--Com SR-3602 \textit{Mini}. Approximate estimates of noise contributions are 25 K from spillover, 50 K from the LNBF, 30 K from the atmosphere, and 5 K sky noise \cite{VSRT77}.
	\begin{figure}[t]
		\centering
		\includegraphics[scale=0.1]{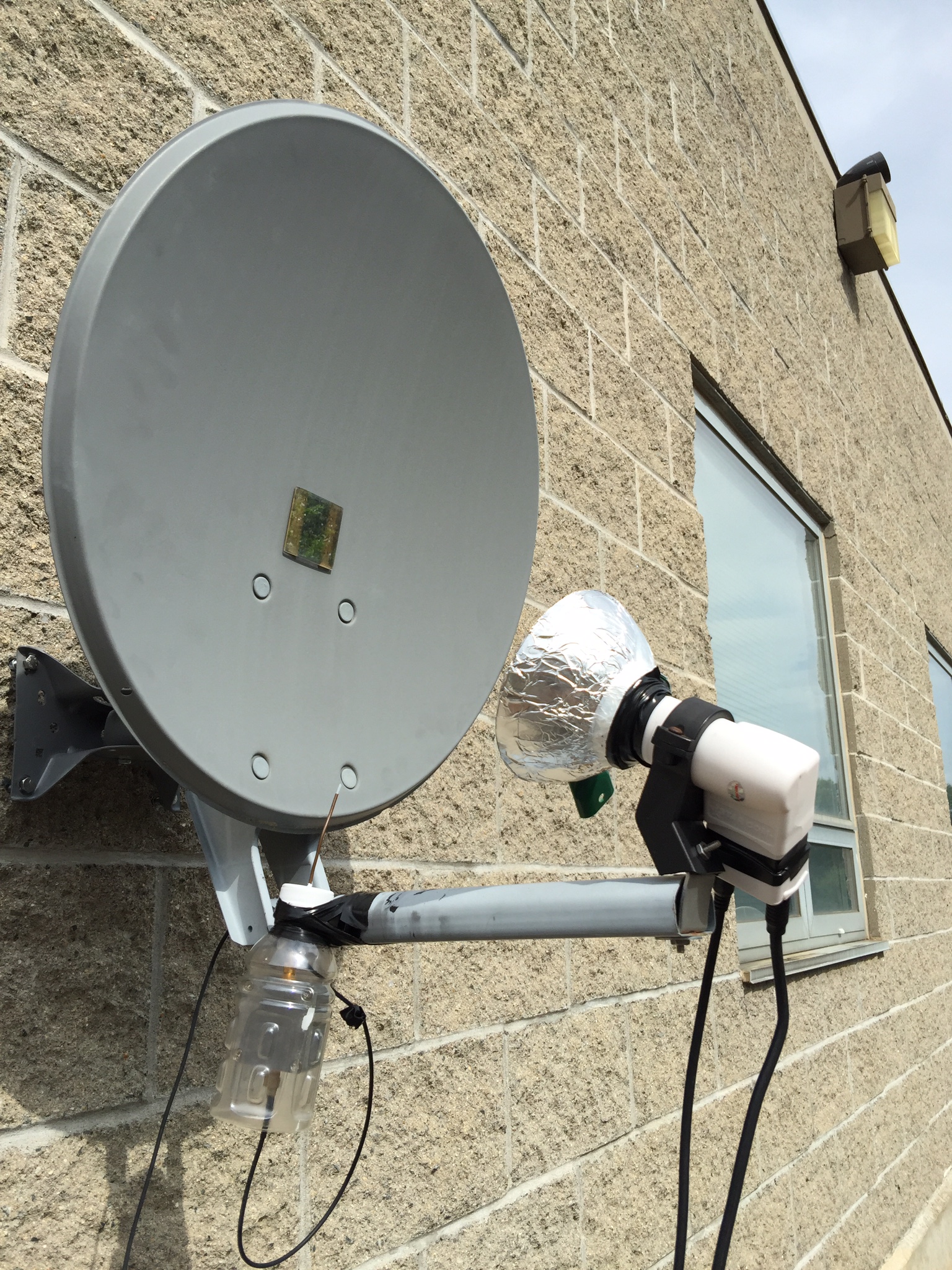}
		\caption{Antenna with LNBF shield.}
		\label{foil}
	\end{figure}		
	
Radiation spillover efficiency was improved to the maximum theoretical value of 91.6\% by positioning the LNBF at an angle of $42^\circ$ towards the dish (or 19.6 cm up from the bottom of the dish; see \textcolor{blue}{Appendix \S\ref{AGEOMAP}} for a more detailed calculation). We note that the efficiency can be boosted to 96\% if the dish were 5.08 cm longer along the circumference, which could be achieved by machining a metal ring and sealing it along the circumference of the dish. We developed an LNBF shield (\textcolor{blue}{Figure} \ref{foil}) in order to further increase the spillover efficiency and lower the clamped back lobe of the LNBF. The shield was constructed from a Norpro 607 plastic casing funnel covered with tinfoil, and tested in the presence of the Haystack radome. It was found to significantly reduce the radar interference, and increase the Y-factor by about 0.1 dB. Note, however, that we do not recommend using the shield for long--term spectrometer deployment, as it may fill with snow and ice in the winter or attract insects and birds. Alternative methods to protect the spectrometer from radio--frequency interference include: (a) mounting the dish on a wall that prevents the LNBF from being in the line of sight of the Clarke belt\footnote{The orbit at 35,786 km that is occupied by geostationary satellites and is a source of radio interference in the X and $\text{K}_\text{u}$ bands.}, (b) extending the size of the dish with ad--hoc metal petals, or (c) inverting the dish and keeping the LNBF shield so that water or snow cannot collect around the tinfoil. Lastly, we note that further software improvements include a high processing efficiency (70\% of the theoretical maximum) with minimum time lost to calibration and calculations. Additionally, the power consumption of the two--channel unit was reduced from 60 W to 15 W due to switching our computer system from a Linux tower to the Intel NUC.	
\section{Single--Channel Velocity Data}
\subsection{Instrumentation}
	\afterpage{\clearpage}

	\begin{figure}[t]
		\centering
		\centerline{\includegraphics[scale=0.3]{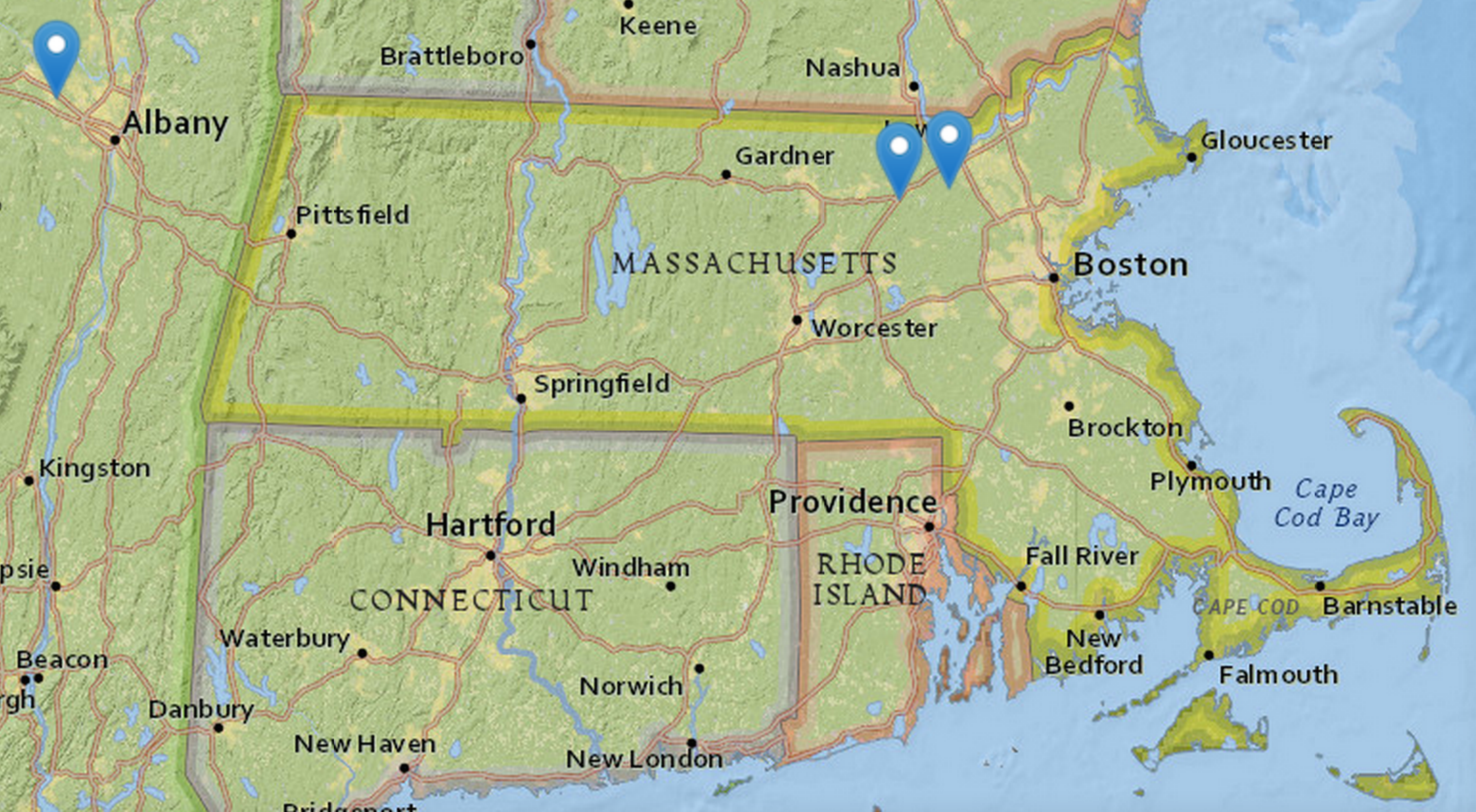}}
		\caption{The geographic network of single--channel ozone spectrometers. The geotags, from left to right, indicate: Union College, MIT Haystack Observatory, and Chelmsford High School. }
		\label{3oz}
	\end{figure}	
	
	\begin{figure}[t]
		\centering
		\includegraphics[scale=0.50]{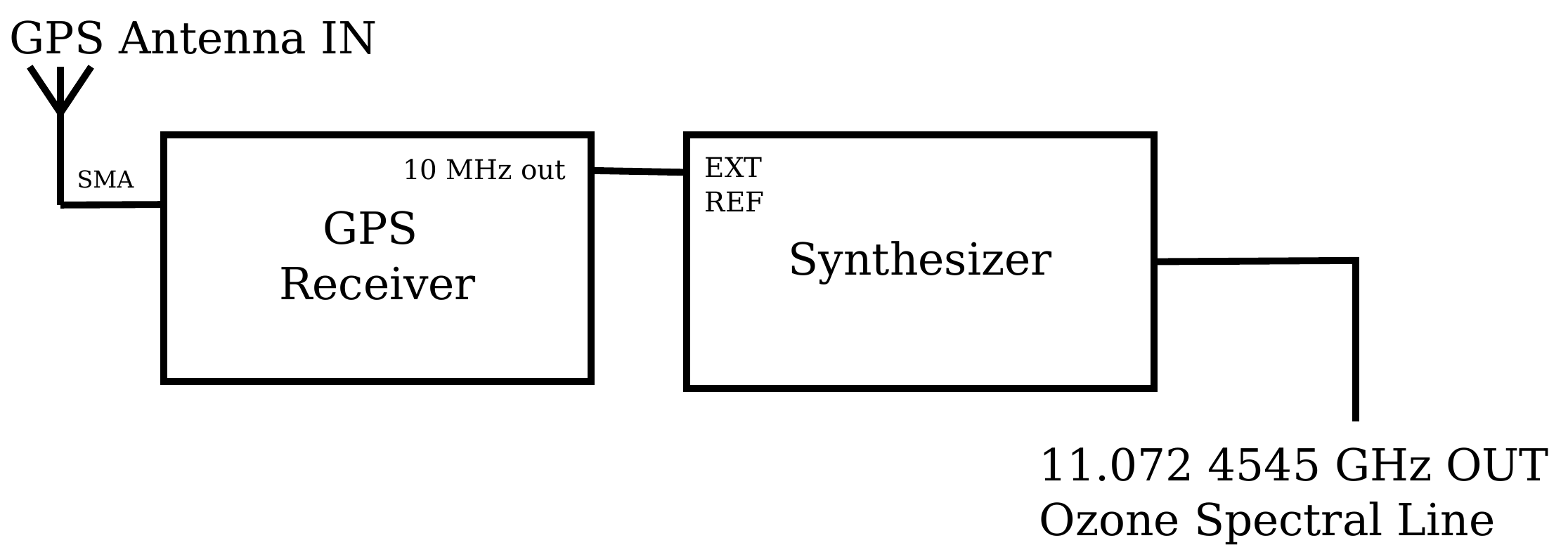}
		\caption{GPS calibration block diagram.}
		\label{gpsref}
	\end{figure}
	
	\begin{figure}[t]
		\centering
		\centerline{\includegraphics[scale=0.7]{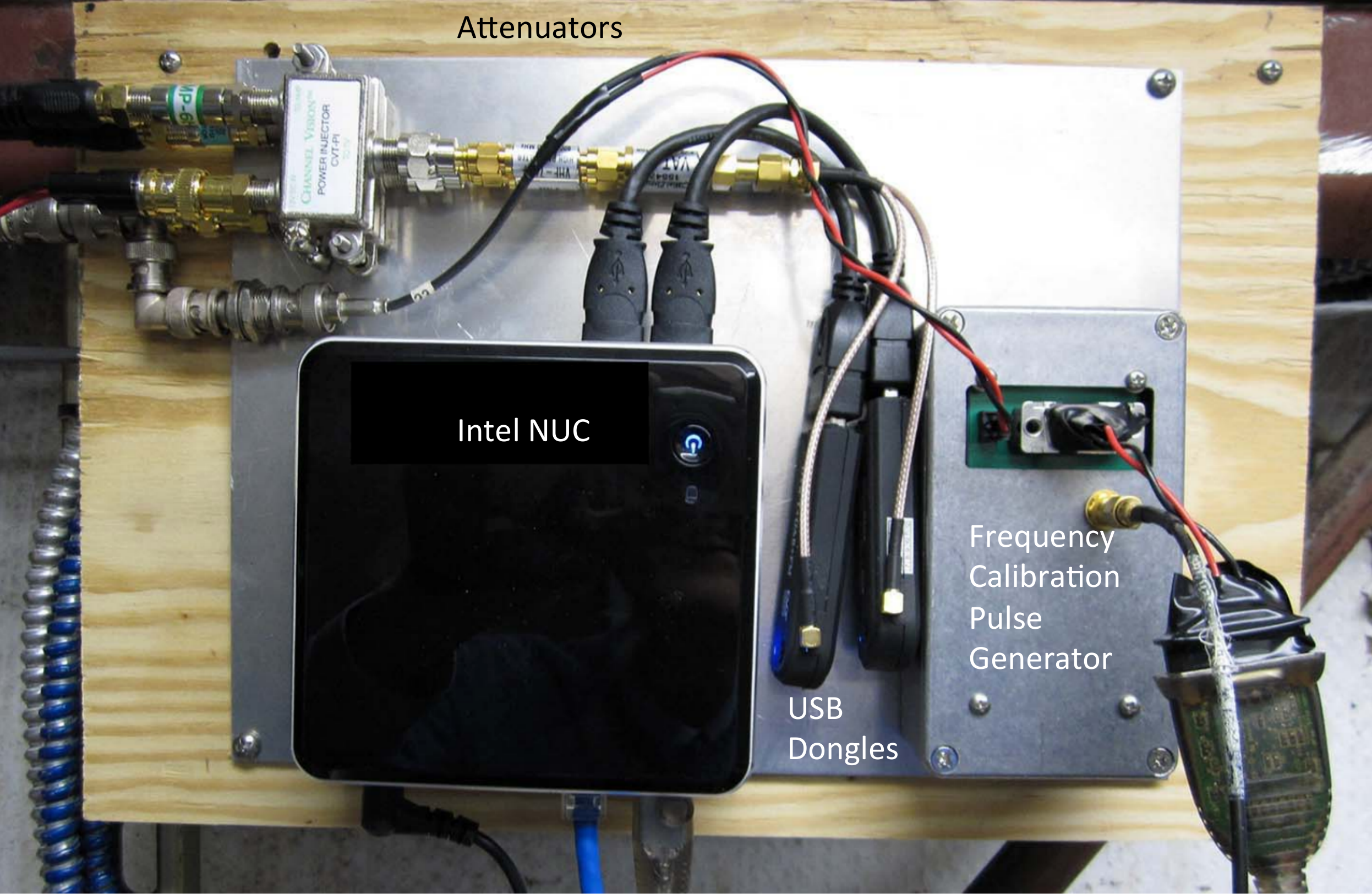}}
		\caption{The physical implementation of the two--channel ozone spectrometer.}
		\label{spectr}
	\end{figure}
	
	\begin{figure}[t]
		\centering
		\centerline{\includegraphics[scale=0.7]{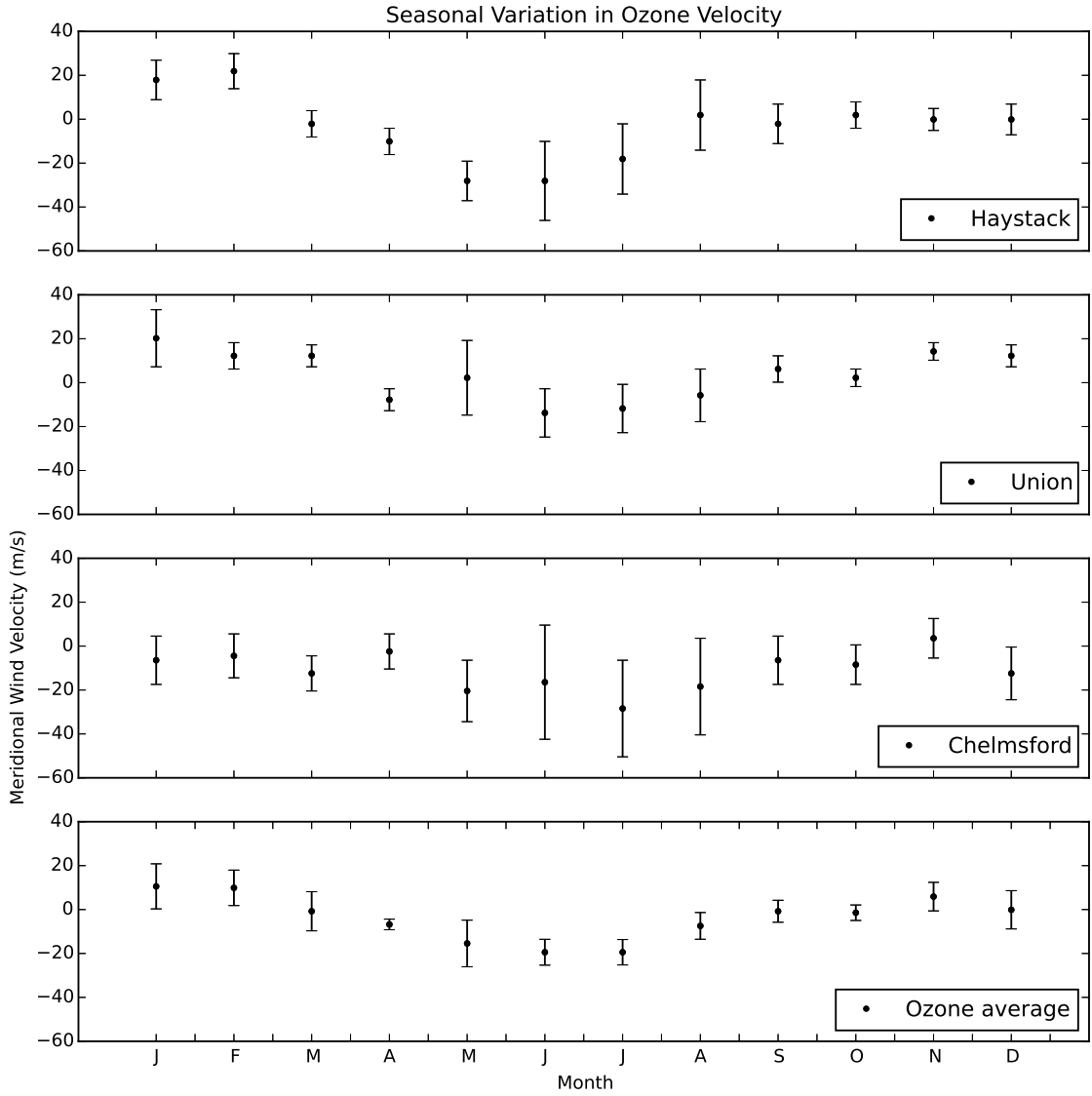}}
		\caption{Seasonal trend in the velocity of ozone at 95 km.}
		\label{ozoneresults}
	\end{figure}	
	
	\begin{figure}[t]
		\centering
		\centerline{\includegraphics[scale=0.7]{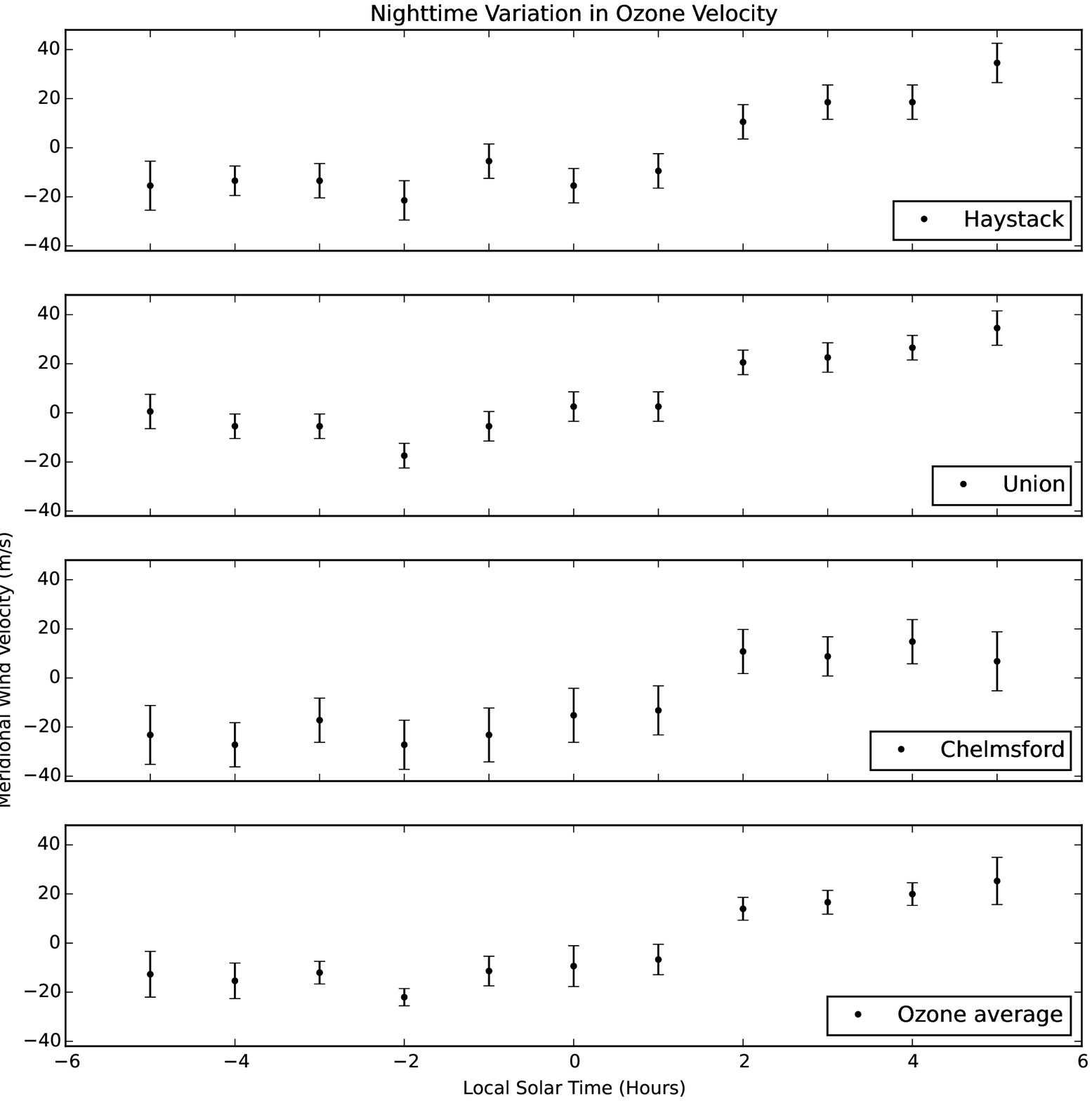}}
		\caption{Nighttime trend in the velocity of ozone at 95 km.}
		\label{ozonelt}
	\end{figure}

To acquire velocity data, we used a network (\textcolor{blue}{Figure} \ref{3oz}) of three single--channel ozone spectrometers located at the MIT Haystack Observatory in Westford, Massachusetts ($42.618^\circ$ N, $71.498^\circ$ W), Union College in Schenectady, New York ($42.817^\circ$ N, $73.928^\circ$ W), and Chelmsford High School in Chelmsford, Massachusetts ($42.619^\circ$ N, $71.367^\circ$ W). The spectrometers are pointed at azimuths $190^\circ$, $180^\circ$, and $172^\circ$, respectively. Note that the beam physically intersects the ozone centered at 95 km above ($37.953^\circ$ N, $71.498^\circ$ W), which is in the Atlantic Ocean about 200 km off the coast of Maryland. Data at Haystack were taken from January 24, 2009 through August 13, 2014. Data at Union College were taken from February 19, 2009 through June 8, 2015. Finally, data at Chelmsford High School were taken from December 25, 2011 through June 8, 2015. The six-channel and two--channel spectrometer readings were not included as we have yet to obtain one full year of data. Before processing the ozone data, we recalibrated the signal reference to the Haystack spectrometer, since its frequency calibration pulse generator circuit uses an atomic clock instead of a crystal oscillator. The atomic clock causes an improper calibration in the local oscillator of USB TV dongles. Using the setup in \textcolor{blue}{Figure} \ref{gpsref}, we first turned the calibration program on, and then initialized the GPS receiver for a stable 10 MHz reference signal, which was used to drive to the synthesizer to produce an output of $11.072~4545$ GHz. We verified that the crystal oscillator has the same effect.

\subsection{Results and Discussion}
\textcolor{blue}{Figure} \ref{ozoneresults} and \textcolor{blue}{Figure} \ref{ozonelt} show velocity data from the spectrometers at Haystack, Union, and Chelmsford. By convention, negative velocities denote winds that travel from north to south (northerly wind), and positive velocities denote winds that travel from south to north (southerly wind). For the general seasonal trend (\textcolor{blue}{Figure} \ref{ozoneresults}), data were obtained by binning over nighttime hours for each day of each year and plotting the weighted arithmetic mean for each month. The weighted standard deviation of the errors in each bin were used to plot the errorbars on each data point. Specifically, for each bin of size $N$ containing velocity data points $v_i$ and error data points $\varepsilon_i,$ the weights are defined to be
	\begin{equation}
		w_i \equiv \frac{1}{\varepsilon_i^2}.
	\end{equation}
The weighted arithmetic mean of the velocities is given by
	\begin{equation}
		v^* = \sum_{i=1}^N v_iw_i
	\end{equation}
and the error in each bin is taken to be the weighted standard deviation, which is the unbiased weighted estimate of the covariance matrix:
	\begin{equation}
		\sigma^* = \left[\frac{\displaystyle\sum_{i=1}^N w_i}{\left(\displaystyle\sum_{i=1}^N w_i\right)^2-\displaystyle\sum_{i=1}^N w_i^2}\right]\displaystyle\sum_{i=1}^N w_i |v_i-v^*|^2.
	\end{equation}
We note that the meridional winds are northerly in the summer and southerly in the winter. The velocity appears to cross the zero mark twice, roughly around April and September, with northerly winds in between, and southerly winds otherwise. The maximum amplitude in the average trend is $30~\text{m}\text{s}^{-1}$, considering a minimum velocity of about --$20~\text{m}\text{s}^{-1}$ in June--July and a maximum velocity of about +$10~\text{m}\text{s}^{-1}$ in January.

A nighttime trend is observed (\textcolor{blue}{Figure} \ref{ozonelt}) in which there are southerly winds from -5 through +1 hours local solar time, and northerly winds from +2 through +5 hours. The transition point appears to be within two hours after midnight, and the full transition is from about --$20~\text{m}\text{s}^{-1}$ to $+20~ \text{m}\text{s}^{-1}$ in amplitude. Daytime data is not available due to the annihilation of ozone by ultraviolet photons from the Sun. Data were binned by day of the year, for each year, and averaged for each hour. The weighted standard deviation in the errors in each bin were again used to plot the errorbars for each data point. The data and trends from all three sites are in good agreement. Additionally, we note that magnitudes in the nighttime trend appear to be slightly larger than the magnitudes in the seasonal trend.
\afterpage{
	\begin{figure*}[t]
		\centering
		\centerline{\includegraphics[scale=0.6]{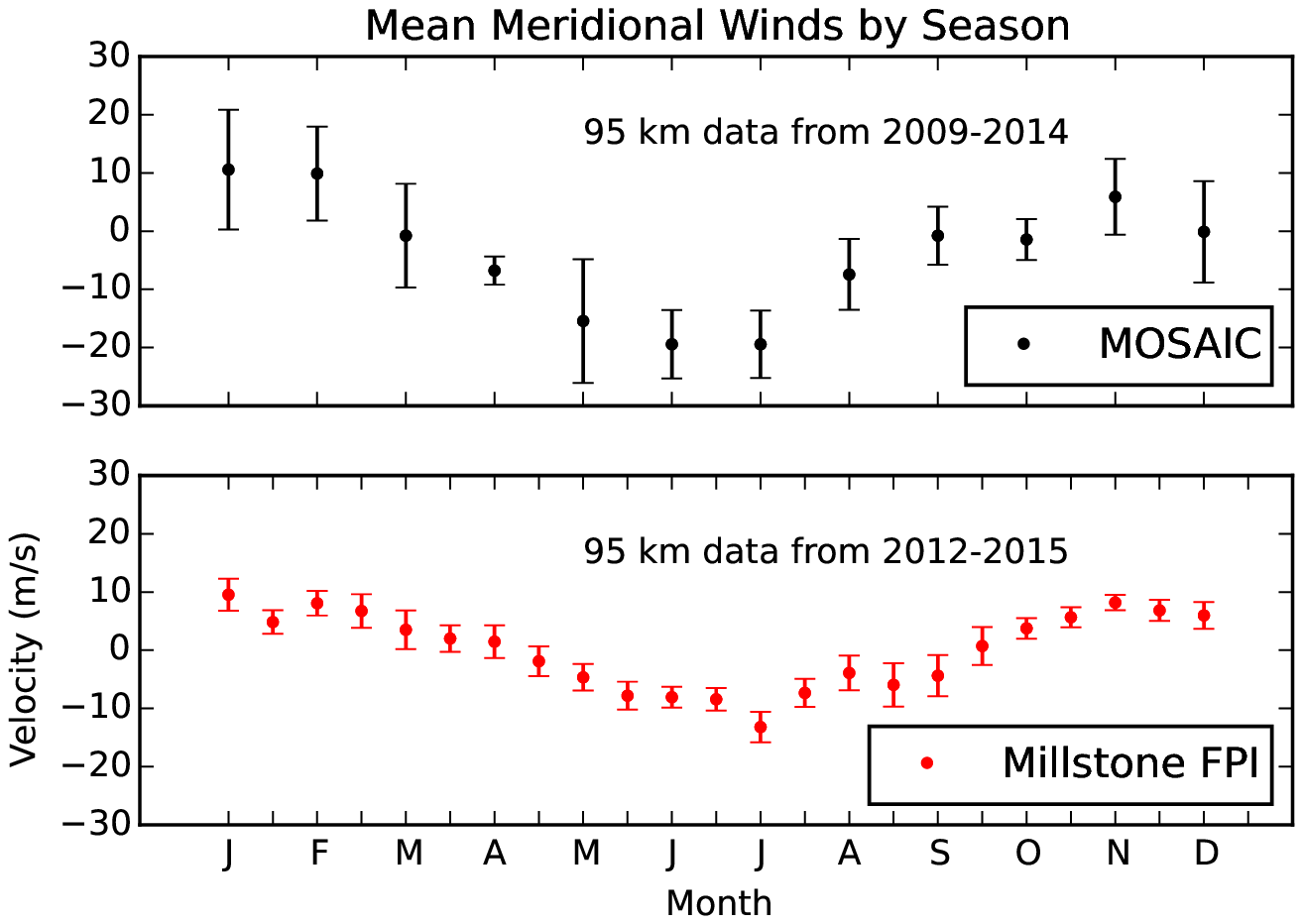}\includegraphics[scale=0.6]{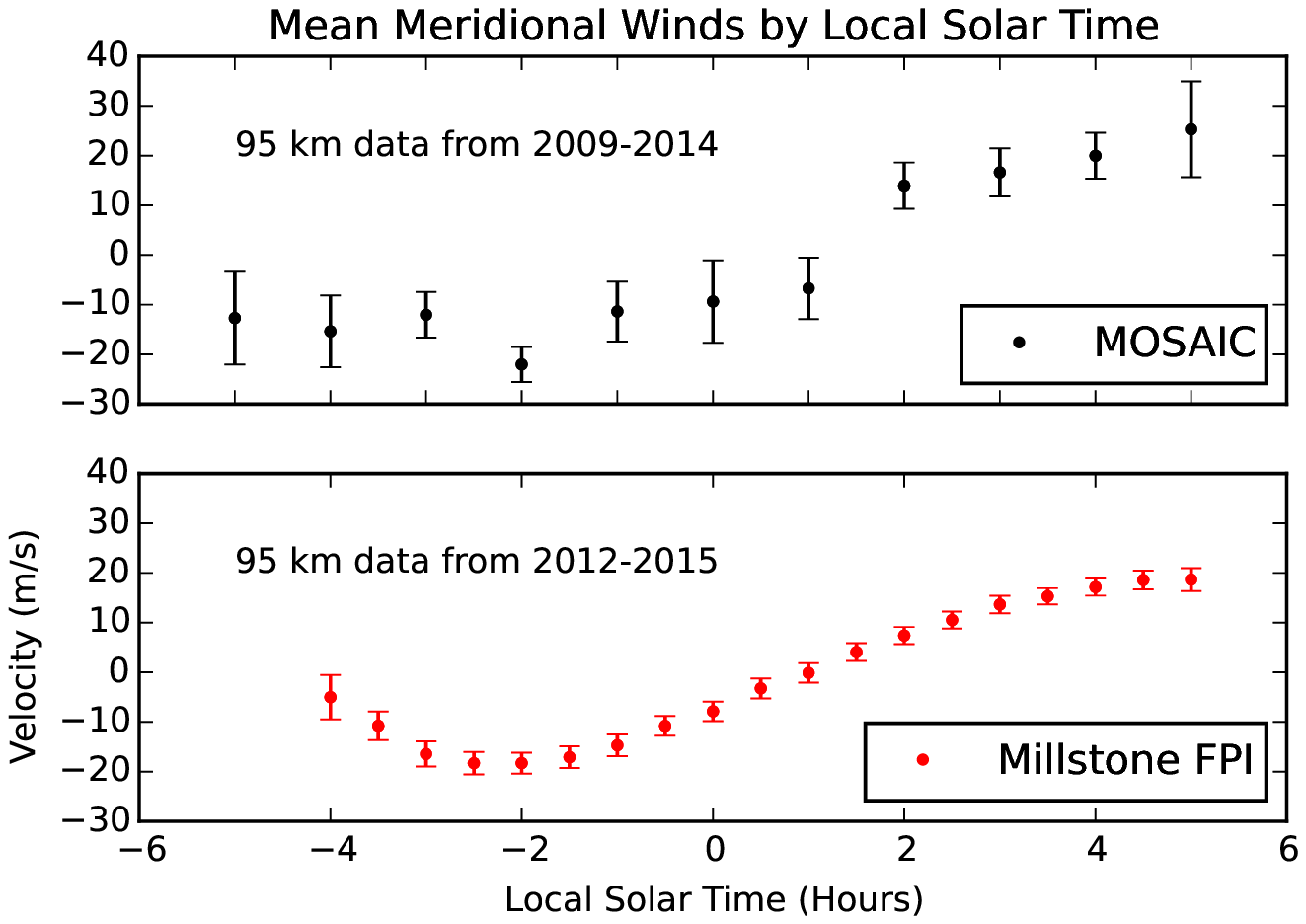}}
		\caption{Comparison of the average meridional velocities of ozone and the oxygen green line by season and local solar time.}
		\label{fpiplot}
	\end{figure*}
	}		
\section{Comparison with Millstone Hill F\'abry--Perot Data}\label{SEC4}
\subsection{Instrumentation}
The Millstone Hill High--Resolution F\'abry--Perot Interferometer (FPI) is located near the MIT Haystack Observatory, specifically at (42.62$^\circ$ N, 71.45$^\circ$ W) where the mean local solar time lags behind universal time (UT) by 4 hours and 46 minutes. The instrument has a 100 mm aperture, and is pressure-tuned. A standard observation at $30^\circ$ elevation involves five optical measurements at azimuths $0^\circ$, $45^\circ$, $135^\circ$, $225^\circ$, and $315^\circ$ \cite{fpi}. One of the measured parameters is the Doppler shift of the 557.7 nm ``green'' spectral line nightglow from atomic oxygen, which is a constituent of the aurora borealis \cite{McLennan1927} and has peak emission at an altitude of 95 km with a half-width half-intensity of 8 km \cite{Gao2012}\cite{Phillips1994}. This makes the altitude of the green line consistent with the altitude of the ozone line at 95 km.
\subsection{Dynamics of the Green Line}
By \textcolor{blue}{Equation} \ref{Dp}, the Doppler shift of the green line produces information about the velocity of atomic oxygen in the upper mesosphere, which can be physically interpreted using winds and tides. The spectral line is the physical result of the $\text{O}$$(^1\text{S}$--$^1\text{D})$ quantum energy level transition in the oxygen atom, and the Barth Mechanism plays the major role in the nighttime green line emission \cite{Gao2012}\cite{Steadman1993}. It is a two-step process:
	\begin{align}
		&\text{O}(^3\text{P})  + \text{O}(^3\text{P}) + \text{M} \rightarrow \text{O}_2^* + \text{M},\\
		&\text{O}(^3\text{P}) + \text{O}_2^* \rightarrow \text{O}(^1\text{S}) + \text{O}_2,
	\end{align}
first involving a three-body collision between two oxygen atoms and an atmospheric chemical $\text{M}$ to produce excited diatomic oxygen, and then a two-body collision to produce the excited state of atomic oxygen.

\subsection{Results}

Due to high optical background illumination from the solar excitation of atomic oxygen \cite{Hedin2009}, the FPI does not take measurements during daytime hours. Scattering from other atmospheric particles and micro--meteors also hampers the ability to acquire precise measurements during the day. Therefore, in our analysis of the FPI data, we only consider the atomic oxygen nightglow between $\pm 5$ hours local solar time. The FPI wind data came from 692 files downloaded from the CEDAR\footnote{Coupling, Energetics, and Dynamics of Atmospheric Regions. The database is located at \url{http://madrigal.haystack.mit.edu/}.} Archival Madrigal Database. The time period was chosen to be 1 January 2012 through 31 December 2015 (automatically pointing to the current day of the year, since it is not yet December). A Python script extracted and processed the data to produce the plots in \textcolor{blue}{Figure} \ref{fpiplot}.

Variations in the green line meridional velocity by season and local solar time appear to correlate well with respective trends in the ozone line. It is observed that the green line winds originate from the north in the summer of the northern hemisphere and from the south in the winter. Furthermore, there is a gradual increase from northerly to southerly winds between -4 and +5 hours local solar time. The uncertainties given by the FPI data are entirely statistical in nature, and are not associated with any flaws in the interferometer itself or the experimental procedure \cite{fpi}. As a whole, the individual error values are low (less than 1.0 $\text{m}\text{s}^{-1}$ on average), resulting in smaller errorbars compared to ozone.
\subsection{Consensus}
The ozone and green line velocity measurements come from two separate instruments, and the fact that they seem to correlate well indicates that these small-scale observations of the middle atmosphere may have some information about the dynamics of global winds. In particular, our results predict transitions in the direction of meridional winds by season and local solar time.

\newpage
\section{Comparison with Other Data and Models}
\subsection{Ascension Island Meteor Radar and Arecibo ISR}
	\afterpage{\clearpage}
	\begin{figure}[t]
		\centering
		\centerline{\includegraphics[scale=0.7]{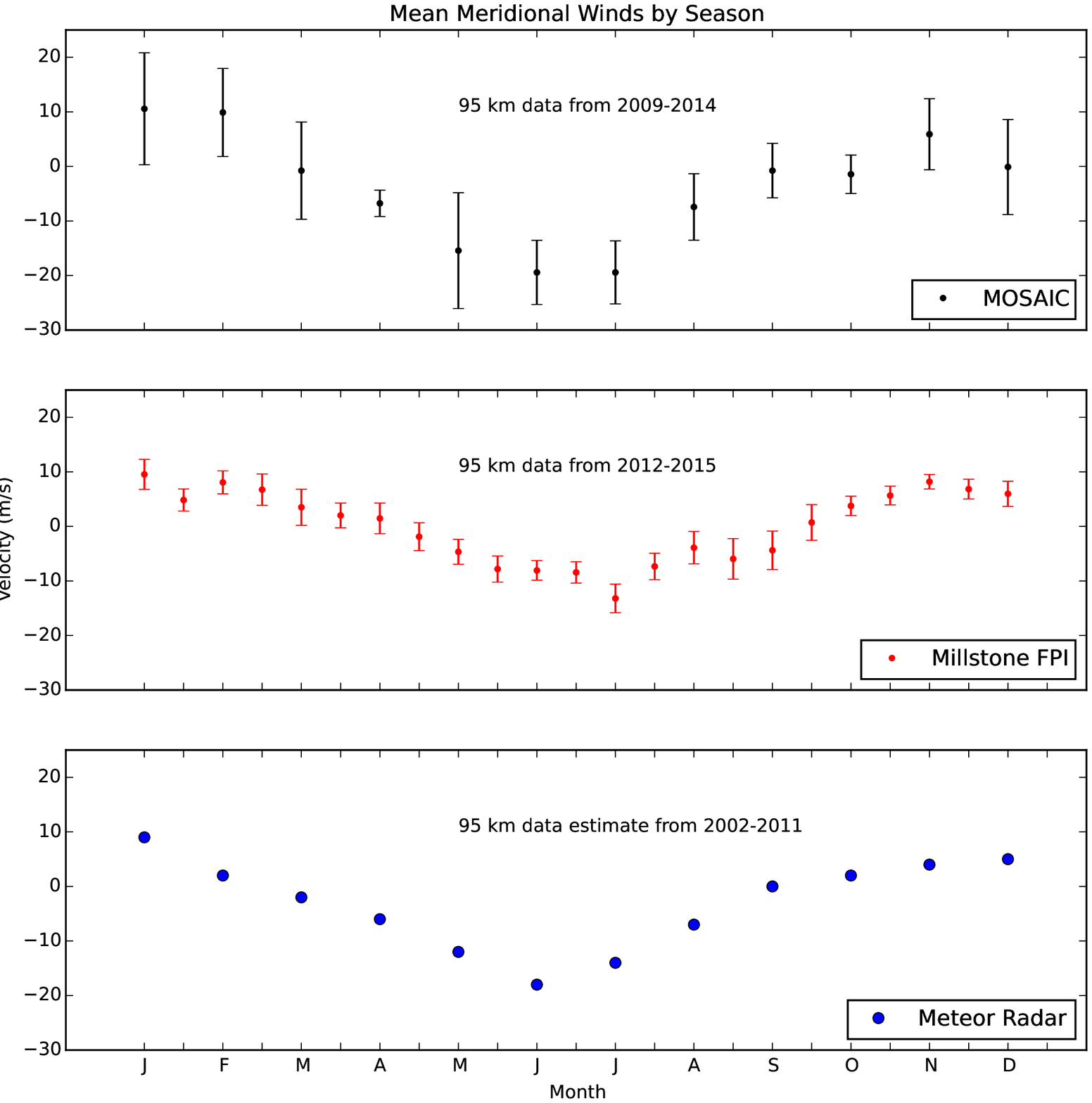}}
		\caption{A comparison of the meridional wind velocities of the ozone, oxygen green line, and atmospheric winds from meteor radar.}
		\label{ascension}
	\end{figure}	
	
	\begin{figure}[t]
		\centering
		\includegraphics[scale=1.2]{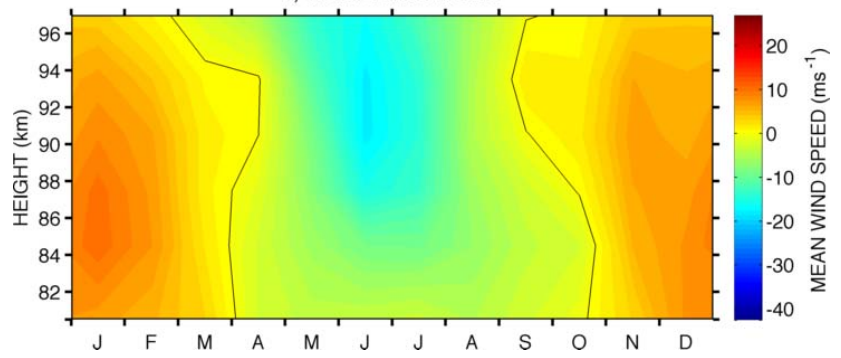}
		\caption{Meridional mean wind from meteor radar measurements over Ascension Island.}
		\label{ascension2}
	\end{figure}	
	
	\begin{figure}[t]
		\centering
		\includegraphics[scale=1.0]{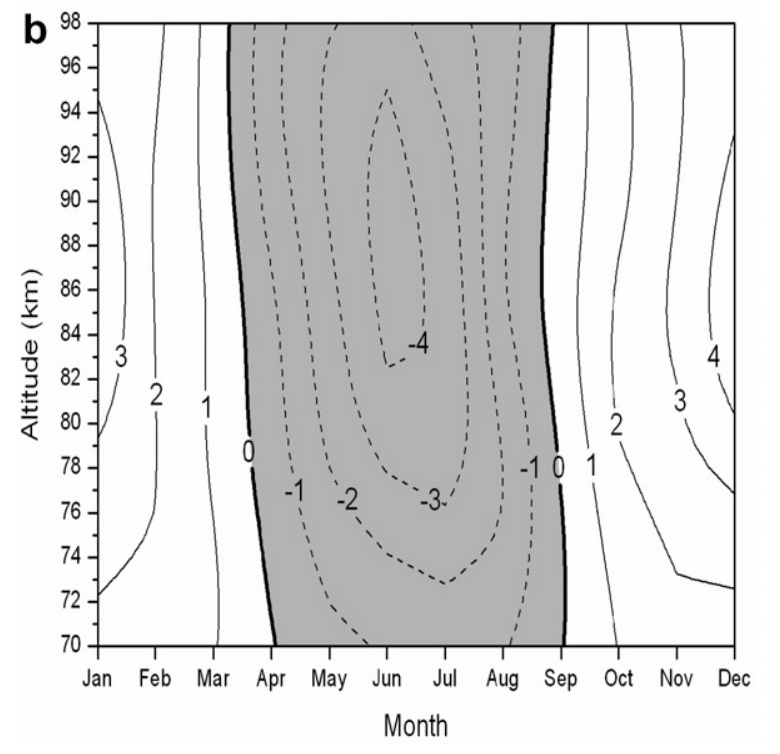}
		\caption{Meridional wind velocities as given by the SOCRATES model.}
		\label{socr}
	\end{figure}
	
	\begin{figure}[t]
		\centering
		\includegraphics[scale=1.20]{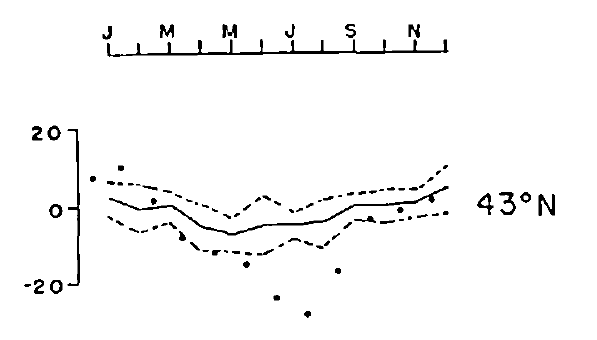}
		\caption{Meridional wind velocities (black dots) as given by the CIRA-72 model.}
		\label{dartt}
	\end{figure}		

We have also compared our data with averaged mesospheric horizontal winds measured over Ascension Island ($8^\circ$ S, $14^\circ$ W) using very high frequency (VHF) meteor radar (MR). The measurements were taken from January 2002 through June 2011 at 95 km \cite{Davis2013}. The results are presented as a color plot (\textcolor{blue}{Figure} \ref{ascension2}) and extracted by season (\textcolor{blue}{Figure} \ref{ascension}). Although the study was undertaken in a tropical latitude, the data clearly show a seasonal correlation in the amplitude of the meridional velocity between the meteor radar, FPI, and ozone data. Additionally, although the following is outside the altitude range of the MOSAIC spectrometer, a similar trend in meridional velocity is seen from Incoherent Scatter Radar (ISR) at the Arecibo Observatory in Puerto Rico targeting the F-region of the ionosphere at 150 km \cite{Kohen2007}.

\subsection{2D--SOCRATES, CIRA--72, HWM, and PWM}
The 2D--SOCRATES model is a theoretical model whose wind velocities are the result of a numerical solution to a partial differential equation with middle-atmosphere boundary conditions. The model is populated with a mix of chemical species, mostly hydrocarbons and chlorofluorocarbons \cite{Xiao2006}\cite{Brasseur2000}. Therefore, there are no oxygen species present in this model to compare with the ozone and green line data. However, the meridional velocities at 95 km (\textcolor{blue}{Figure} \ref{socr}) have a similar trend in season, but with smaller magnitudes. This may be the result of zonally-averaged residual meridional wind calculations, or atmospheric tide extraction over all latitudes and longitudes, which would decrease the overall magnitudes. A similar trend in meridional velocity by season is also seen in the CIRA-72 wind model (\textcolor{blue}{Figure} \ref{dartt}) near 90 km altitude at $43^\circ$ N altitude \cite{Dartt1983}. 

The Horizontal Wind Model (HWM) is an empirical model based on zonal and meridional wind data from numerous studies of the atmosphere under a range of climatologically averaged geophysical conditions \cite{Hedin1996}\cite{Drob2008}. The original 1990 model, HWM90, was revised three times to produce HWM93, HWM07, and HWM15 in 1993, 2007, and 2015, respectively. While experimental data from certain regions of the atmosphere (notably the troposphere) seem to match well with the model, the team that produced HWM93 directly indicated significant discrepancies regarding wind data near the mesopause \cite{Hedin1996}. Other studies have cited inconsistencies between HWM and experimental observations, notably FPI measurements \cite{Sandford2010}\cite{Yuan2013}. 

Finally, we considered the 2D prevailing wind model, which incorporates climatologically averaged data from 46 ground-based medium-frequency (MF) radar and meteor radar (MR) stations, as well as the HRDI satellite \cite{Portnyagin2004}. The seasonal trend in the meridional wind velocity in prevailing wind model did not correlate with data from ozone, Millstone Hill, or Ascension Island.\newpage
\section{Preliminary Observations of Ozone Temperature}

In previous work, the theoretical ozone temperature in the mesosphere and lower thermosphere (MLT) was determined to be around 170 K due to particle collisions and ultraviolet photon absorption. Measurements of the ozone line by initial Very Small Radio Telescope (VSRT) hardware had insufficient spectral resolution to provide an accurate measurement of the ozone temperature, but did suggest that the temperature was below $180$ K \cite{VSRT43}. Recent nighttime data taken from the network of three ozone spectrometers (Haystack, Chelmsford, Union) from 20 January 2015 through 10 May 2015 (winter and spring) reveal an average ozone temperature of $170\pm12$ K. However, looking at one individual measurement from only the Haystack spectrometer reveals an ozone temperature of $120\pm118$ K. Thus, singular measurements of the ozone temperature seem to have erratically high errorbars, while the average has a lower overall temperature. By comparison, satellite observations of the oxygen green line show temperatures in the $180$--$200$ K range at 95 km altitude and $45^\circ$ N latitude \cite{Gao2012}. Further observations by Rayleigh--Mie--Raman lidar and potassium resonance lidar show a seasonal temperature variation of 160--190 K at 95 km \cite{Gerding2008}. Lastly, sodium lidar measurements in Fort Collins, Colorado ($41^\circ$ N, $105^\circ$ W) show a seasonal temperature variation of 180-190 K at 95 km \cite{Yuan2008}.

\section{Future Work}	
\subsection{The Global MOSAIC Network}
In contrast to satellite-based instruments, ground-based spectrometers have the advantage of time resolution but are weak in spatial resolution \cite{Portnyagin2004}. In order to achieve a greater spatial coverage in latitude and longitude, M. J. Kosch of Lancaster University has proposed to expand the MOSAIC network to encompass eighteen two--channel spectrometers that will be located across the United States of America (\textcolor{blue}{Table} \ref{AmNet}), Europe (\textcolor{blue}{Table} \ref{EurNet}), Africa (\textcolor{blue}{Table} \ref{AfNet}), and Antarctica (\textcolor{blue}{Table} \ref{AntNet}). Potentially, some of these spectrometers would also be pointed toward the east or west in order to obtain data for the zonal wind velocity of ozone at 95 km. This would provide a complete map of the two-dimensional motion of the ozone, and also reveal more about the global wind dynamics at that altitude. Increased averaging in the data (by increasing the number of instruments) would also fine-tune the resolution of ozone temperature.

\begin{table}[]
\centering
\caption{The American MOSAIC Network}
\label{AmNet}
\centerline{\begin{tabular}{@{}lllll@{}}
\toprule
\multicolumn{1}{c}{{\bf Spectrometer Location}} & \multicolumn{1}{c}{{\bf Town}} & \multicolumn{1}{c}{{\bf State}} & \multicolumn{1}{c}{{\bf Latitude ($^\circ$N)}} & \multicolumn{1}{c}{{\bf Longitude ($^\circ$E)}} \\ \midrule
Chelmsford High School                          & Chelmsford                     & Massachusetts                   & 42.62                                 & 288.63                                 \\
MIT Haystack Observatory                        & Westford                       & Massachusetts                   & 42.50                                 & 288.53                                 \\
Bridgewater State University                    & Bridgewater                    & Massachusetts                   & 41.99                                 & 288.99                                 \\
Lynnfield High School                           & Lynnfield                      & Massachusetts                   & 42.54                                 & 288.97                                 \\
Union College                                   & Schnectady                     & New York                        & 42.80                                 & 286.07                                 \\
University of North Carolina                    & Greensboro                     & North Carolina                  & 36.07                                 & 280.17                                 \\
Alaska Pacific University                       & Anchorage                      & Alaska                          & 61.19                                 & 210.20                                 \\ \bottomrule
\end{tabular}}
\end{table}

\begin{table}[]
\centering
\caption{The European MOSAIC Network}
\label{EurNet}
\centerline{\begin{tabular}{@{}lllll@{}}
\toprule
\multicolumn{1}{c}{{\bf Spectrometer Location}} & \multicolumn{1}{c}{{\bf Region}} & \multicolumn{1}{c}{{\bf Country}} & \multicolumn{1}{c}{{\bf Latitude ($^\circ$N)}} & \multicolumn{1}{c}{{\bf Longitude ($^\circ$E)}} \\ \midrule
Ny-\r{A}lesund                                      & Svalbard                         & Norway                            & 78.93                                 & 11.92                                  \\
Sodankyl\"{a}                                      & Lapland                          & Finland                           & 67.42                                 & 26.58                                  \\
University of Lancaster                         & Lancashire                        & United Kingdom                    & 54.05                                 & 357.20                                 \\ \bottomrule
\end{tabular}}
\end{table}

\begin{table}[]
\centering
\caption{The African MOSAIC Network}
\label{AfNet}
\centerline{\begin{tabular}{@{}lllll@{}}
\toprule
\multicolumn{1}{c}{{\bf Spectrometer Location}} & \multicolumn{1}{c}{{\bf Region}} & \multicolumn{1}{c}{{\bf Country}} & \multicolumn{1}{c}{{\bf Latitude ($^\circ$N)}} & \multicolumn{1}{c}{{\bf Longitude ($^\circ$E)}} \\ \midrule
Cairo                                           & Cairo Governorate                & Egypt                             & 30.05                                  & 31.23                                  \\
Bahir Dar                                       & Amhara                           & Ethiopia                          & 11.60                                  & 37.38                                  \\
Nairobi                                         & Nairobi County                   & Kenya                             & 358.72                                 & 36.82                                  \\
Windhoek                                        & Khomas                           & Namibia                           & 337.43                                 & 17.08                                  \\
Sutherland                                      & Northern Cape                    & South Africa                      & 327.61                                 & 20.66                                  \\ \bottomrule
\end{tabular}}
\end{table}

\begin{table}[]
\centering
\caption{The Antarctic MOSAIC Network}
\label{AntNet}
\centerline{\begin{tabular}{@{}lllll@{}}
\toprule
\multicolumn{1}{c}{{\bf Spectrometer Location}} & {\bf Region}    & \multicolumn{1}{c}{{\bf Continent}} & \multicolumn{1}{c}{{\bf Latitude ($^\circ$N)}} & \multicolumn{1}{c}{{\bf Long ($^\circ$E)}} \\ \midrule
Halley Bay (UK)                                 & Brunt Ice Shelf & Antarctica                          & 284.42                                 & 333.43                                 \\
SANAE IV (South Africa)                         & Vesleskarvet    & Antarctica                          & 288.33                                 & 357.16                                 \\
Amundsen--Scott Station (USA)                   & South Pole      & Antarctica                          & 270.00                                 & 0.00                                   \\ \bottomrule
\end{tabular}}
\end{table}

\subsection{Further hardware upgrades}
Ideally, the Global MOSAIC Network would make use of the atomic clock--based frequency calibration pulse generator, instead of the one that is based on the crystal oscillator. The MIT Haystack Observatory has placed an order for several new atomic clocks that will eventually be sent to M. J. Kosch for field deployment.

\subsection{Helping to resolve the theory of pole--to--pole flow}
	\begin{figure}[h!]
		\centering
		\includegraphics[width=0.60\textwidth]{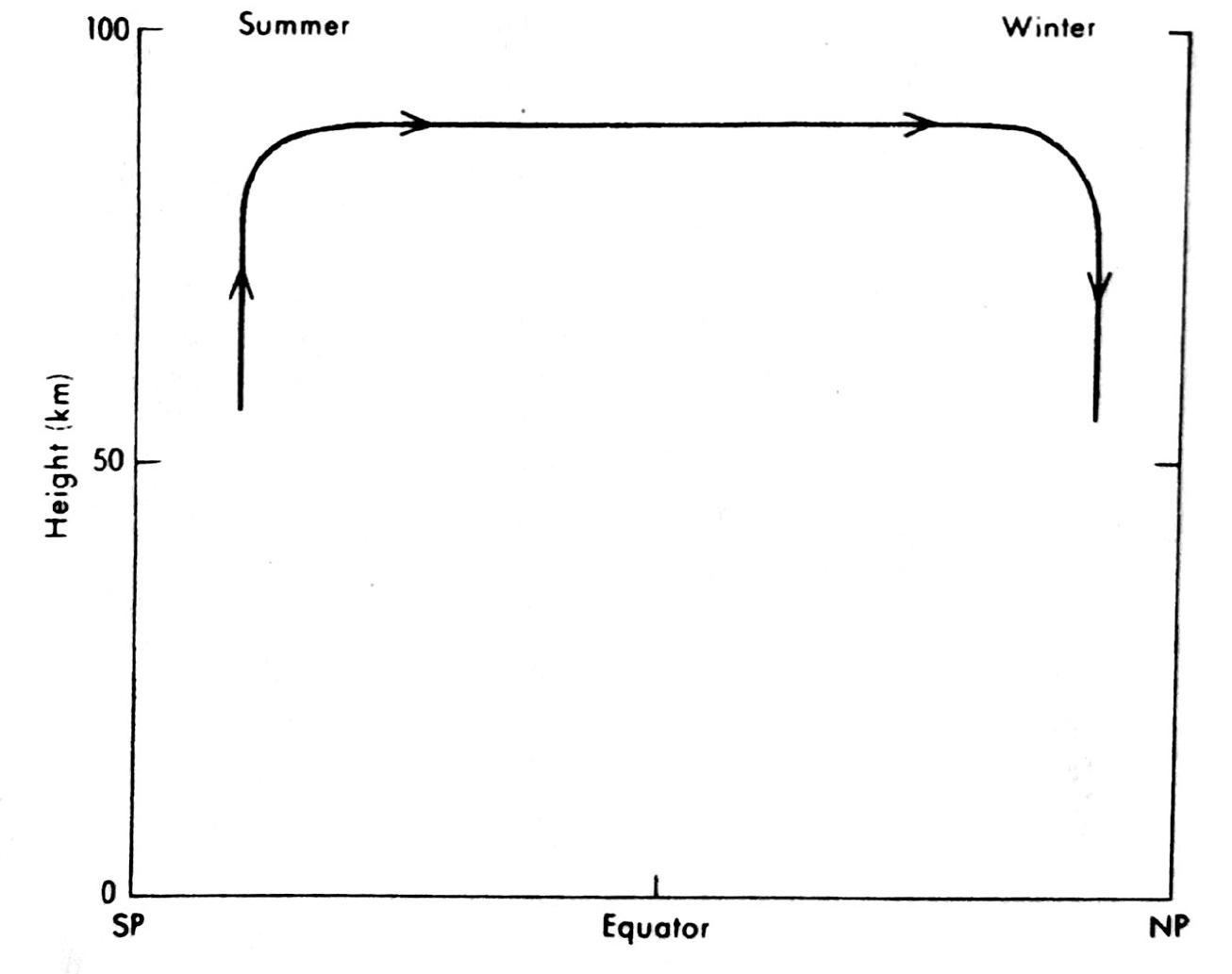}
		\caption{Inter--Hemispheric Circulation.}
		\label{ppf}
	\end{figure}
\noindent Lastly, we note that there exists a theory of inter--hemispheric wind circulation in the mesosphere, also known as pole--to--pole flow \cite{PhysAtmos}. As shown in \textcolor{blue}{Figure} \ref{ppf}, there is a seasonal trend in the mesospheric wind, in which air from the south pole moves upward through the lower thermosphere in the summer, and remains in the lower thermosphere until the winter, when it moves back into the mesosphere. This may be related in some form to the seasonal change in the polarity of the meridional winds as observed by the single--channel ozone spectrometers, FPI, and meteor radar. Earlier work by Yuan indicated a summer--to--winter pole--to--pole meridional flow of 17 $\text{m}\text{s}^{-1}$ at 86 km \cite{Yuan2008}, which is within the range of magnitudes that we have observed at 95 km.\newline

\noindent Using the Global MOSAIC Network, further studies of the meridional and zonal velocities of ozone could reveal novel information about the Earth's pole--to--pole circulation as well as the dynamics of air flow between the MLT and other regions of Earth's atmosphere. In conclusion, we hope to see the two--channel hardware replicated and deployed at key locations around the world in the coming years in order to realize a global network of spectrometers that will provide higher-resolution data about the dynamics of ozone in the middle atmosphere.

\section{Appendix: Optimizing Antenna Geometry}\label{AGEOMAP}
\subsection{Summary}
Here, we develop a theoretical model for the antenna geometry, which is incorporated into the physical implementation of the two--channel ozone spectrometer. A set of equations are developed that relate the angle of the low-noise block downconverter feedhorn antenna (LNBF) to the radiation flux that hits the parabolic reflector. A simulation in Python determines that the spillover efficiency is maximized, and hence power loss is minimized, when the LNBF is pointed towards the dish at an angle of 41 degrees.
\subsection{Model}
	\begin{figure}[h!]
		\centering
		\includegraphics[width=0.40\textwidth]{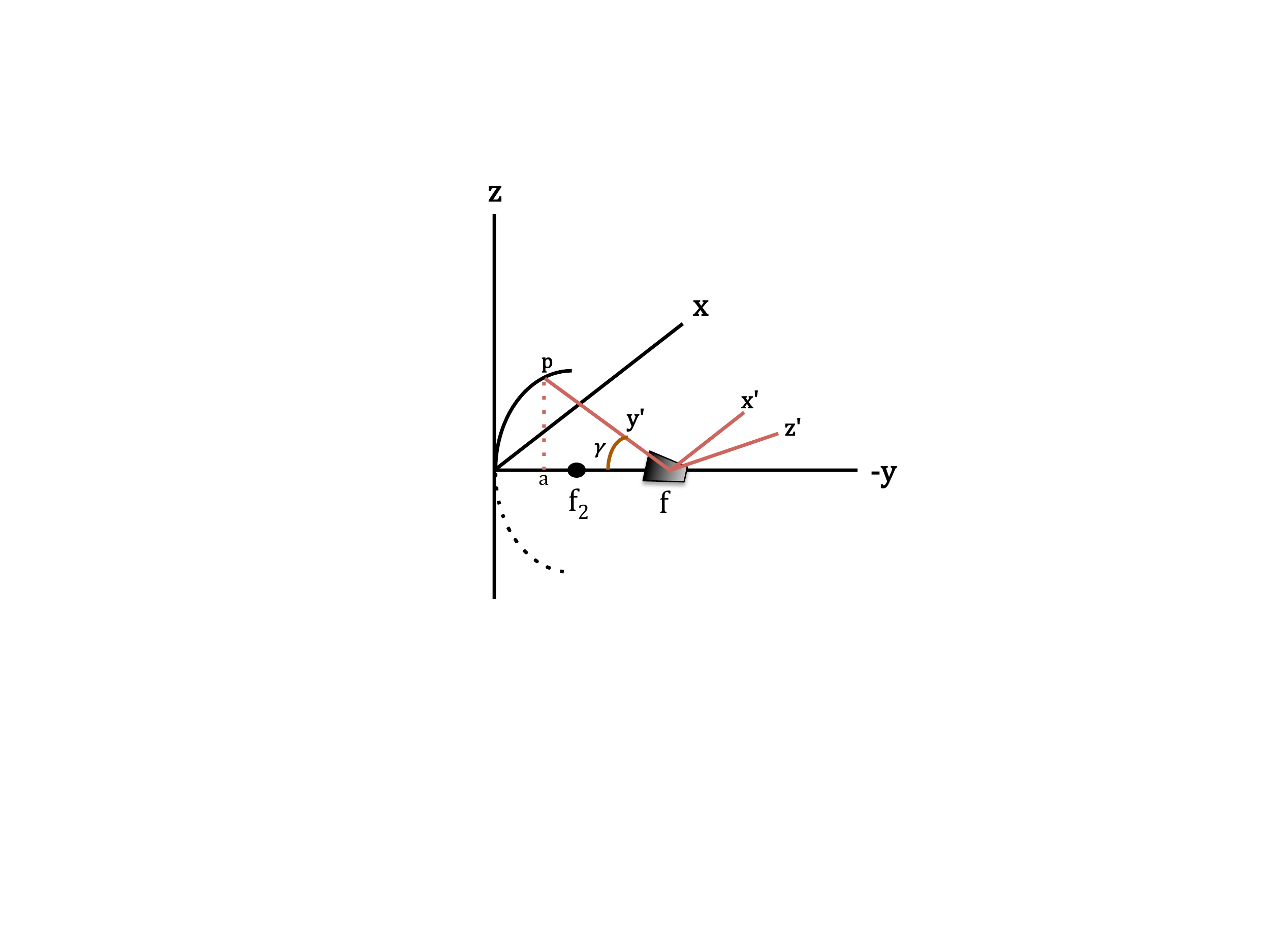}
		\caption{Parabolic antenna geometry.}
		\label{A}
	\end{figure}
	
In \textcolor{blue}{Figure} \ref{A}, we model the {Winegard DS-4040 dish} as an offset parabola with $0.4572$ m diameter that extends into a region spanned by the $\{\bold{\hat{x}},\bold{\hat{y}},\bold{\hat{z}}\}$ basis with origin $\mathcal{O}_1$ and focus $f_2 = 10.0$ cm. The dish itself is the solid black line in the positive octant, while the full parabola includes the dashed line extending into the negative octant. The unit vector $\bold{\hat{x}}$ represents \textcolor{black}{East}, $\bold{\hat{y}}$ represents \textcolor{black}{North}, and $\bold{\hat{z}}$ represents \textcolor{black}{Up}. In unit spherical coordinates, the location of an arbitrary point in this basis is given by
	\begin{equation}
\left\{\begin{matrix}
&x=\cos\theta\cos\varphi\\ 
&y=\cos\theta\sin\varphi\\ 
&z=\sin\theta
\end{matrix}\right.	\end{equation}
where $\theta$ and $\varphi$ are the altitudinal and azimuthal angles, respectively. The antenna feed is located at origin $\mathcal{O}_2$ with focus $f=10.6$ cm\footnote{The beam focus center was measured to be 0.6 cm into the circular waveguide of the LNBF, and not entirely at the ``dot" on the cover of the feed.}, and any vector emerging from this origin is described by the ``primed" basis, which consists of the unit vectors $\{\bold{\hat{x}'},\bold{\hat{y}'},\bold{\hat{z}'}\}$. As in \textcolor{blue}{Figure} \ref{A}, to rotate the antenna by some angle $\gamma$, we choose to keep the $\bold{\hat{x}'}$-axis constant and then apply a rotation matrix on the $yz$-plane:
	\begin{equation}
		\begin{bmatrix}
y'\\ 
z'
\end{bmatrix} = \begin{pmatrix}
\cos\gamma &\sin\gamma \\ 
 -\sin\gamma&\cos\gamma 
\end{pmatrix}\begin{bmatrix}
y\\ 
z
\end{bmatrix}.
	\end{equation}
This results in the following transformed coordinate values:
\begin{equation}
\left\{\begin{matrix}
&x'=x\\ 
&y'=y\cos\gamma+z\sin\gamma\\ 
&z'=z\cos\gamma-y\sin\gamma.
\end{matrix}\right.
\end{equation}
The transformed LNB angle $\gamma'$ is:
	\begin{equation}
		\gamma'=\arctan\left[\frac{\left(x'^2+z'^2\right)^{1/2}}{y'}\right].
	\end{equation}
The location of some point $p$ is described by the parabola equation
	\begin{equation}
		p^2 = x^2 + z^2 = 4fa,
	\end{equation}
where $a$ is the projection of $p$ onto the negative $y$-axis. 
	\begin{figure}[h]
		\centering
		\includegraphics[width=0.40\textwidth]{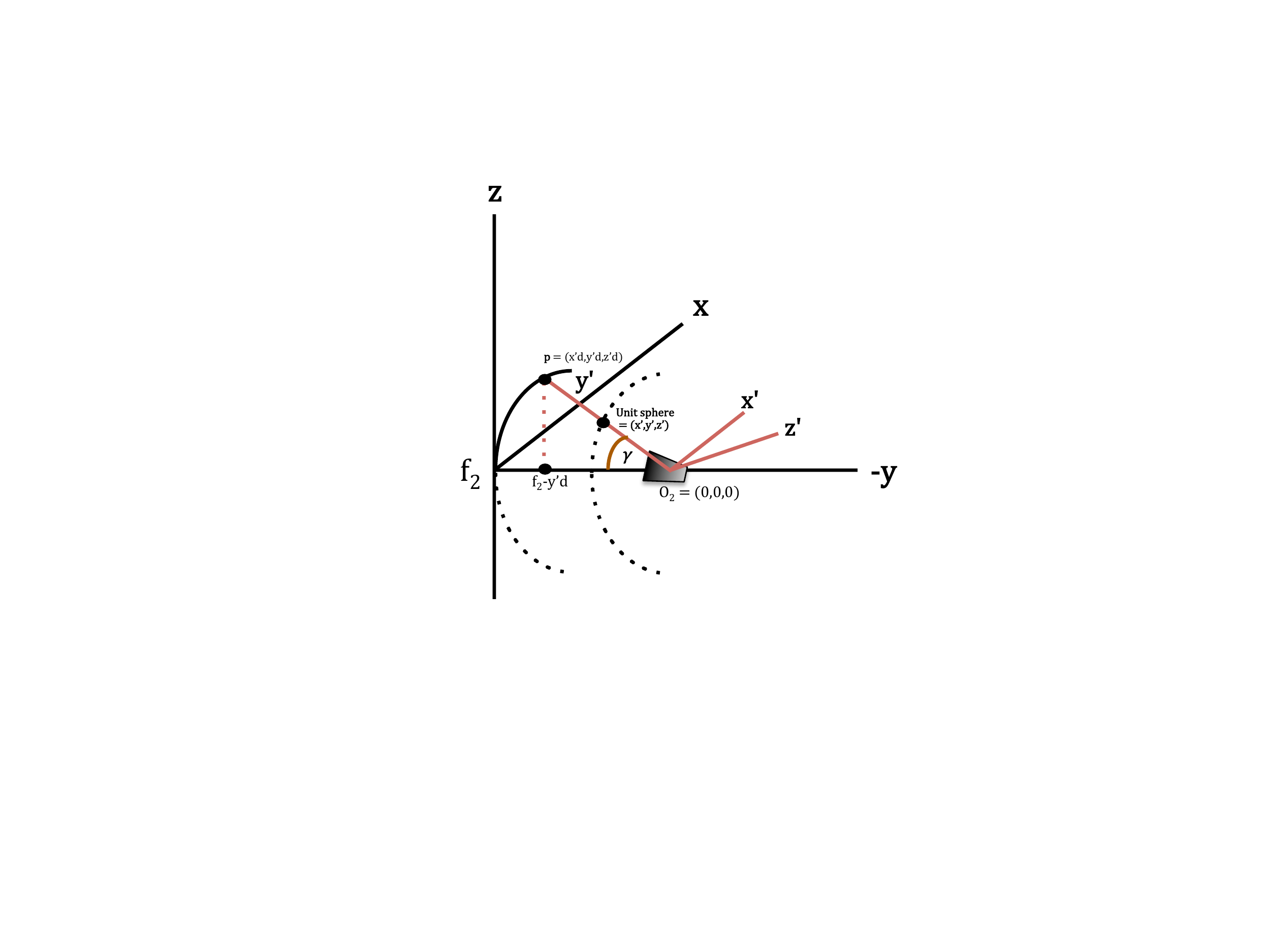}
		\caption{Coordinate scaling.}
		\label{C}
	\end{figure}
However, since we want to look at the situation from the perspective of the antenna, we must shift over to origin $\mathcal{O}_2$. As depicted in \textcolor{blue}{Figure} \ref{C}, we take the unit coordinates emerging from $\mathcal{O}_2$ and choose some scaling factor $d$ that makes $\bold{p}=(x'd)\,\bold{\hat{x}'}+(y'd)\,\bold{\hat{y}'}+(z'd)\,\bold{\hat{z}'}.$  Consequently, 
\begin{equation}
f_2-y'd = a,
\end{equation}
and the parabola equation is modified accordingly:
	\begin{equation}
		(x'^2+y'^2)d^2=4f(f_2-y'd).
	\end{equation}
We realize that this is a quadratic equation in $d$ with coefficients
\begin{equation}
	\left\{\begin{matrix}
&\alpha=(x'^2+z'^2)/(4f)\\ 
&b=y\\ 
&c=-f_2
\end{matrix}\right.
\end{equation}
and solution
	\begin{equation}
		d = \frac{-b+\sqrt{b^2-4\alpha c}}{2\alpha}
	\end{equation}
where we only consider the positive square root as it is physically realizable. The offset parabola, in the scaled coordinate system, is now extended into circle with radius $r$ and a $y$-offset of $r$ to account for the antenna actually being shifted by that amount. The quantity $R$ is defined to be
\begin{equation}
	R \equiv \sqrt{\frac{(y'd-r)^2+(x'd)^2}{r^2}}
\end{equation}
and is used as follows: if $R$ = 1, then the beam emanating from the antenna is at the edge of the dish. If $R<1,$ then the beam spills over past the edge (\textit{i.e.} it is not on the circle), resulting in a power loss because not all of the radiation is striking the dish. Quantitatively, the \textit{spillover efficiency} $\epsilon_s$ is the ratio of flux intensities \cite{Balanis1997} given by
	\begin{equation}\label{integrals}
		\epsilon_s = \frac{\displaystyle\iint\limits_{\{r<1\}}\mathrm{d}\Omega\,\,G(\theta,\varphi,\gamma)\cos\theta}{\displaystyle\oiint\,\mathrm{d}\Omega\,\,G(\theta,\varphi,\gamma)\cos\theta}
	\end{equation}
where $\Omega$ is the solid angle subtending the dish (for all $r<1$ in the numerator, and across the entire region in the denominator) and $G(\theta,\varphi,\gamma)$ is the gain function of the antenna, which we define as a Gaussian:
	\begin{align}
		G(\theta,\varphi,\gamma) &= \exp\left(-k\gamma'^2\right) \\
		&= \exp\left(-k\arctan f\right)
	\end{align}
where 
\begin{equation}
	f\equiv f(\theta,\varphi,\gamma) = \frac{(\cos\theta)\sqrt{\cos^2\varphi+\sin^2\varphi\sin^2\gamma}}{\cos\gamma\cos\theta\sin\varphi+\sin\gamma\sin\theta}.
\end{equation}
The quantity $\arctan f$ is the antenna angle in the $\mathcal{O}_2$ basis and $k=-0.692/(22)^2$ is a normalization constant that takes into account the half-width half-maximum (HWHM) power at 22 degrees orientation.
\subsection{Results}
\begin{figure}[t]
		\centering
	\includegraphics[scale=0.6]{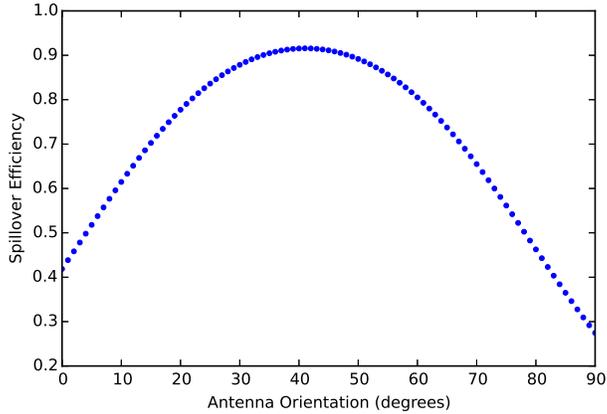}
		\caption{Theoretical spillover efficiency versus antenna angle.}
		\label{LNBTEST}
	\end{figure}

Due to a lack of efficient computational resources, direct symbolic integration of \textcolor{blue}{Equation} \ref{integrals} is not feasible with Wolfram Mathematica, so we perform numerical integration in Python. A plot of the output of the program is shown in \textcolor{blue}{Figure} \ref{LNBTEST}. We find that a maximum efficiency of $0.916$ occurs when the antenna is pointed at $\gamma=41$ degrees from the horizontal. 
\section{Appendix: LNBF Characterization}\label{LNBAP}
\subsection{Introduction}
The \textit{Y-factor} is widely used to measure the noise temperature of an amplifier. It is defined to be the ratio of the amplifier power when ``hot" to the power when ``cold,'' and mathematically, this is represented by the following equation:
	\begin{equation}\label{AOK}
		Y = \frac{P_\text{hot}}{P_\text{cold}} = \frac{T_\text{amb}+T_\text{LNBF}}{T_\text{sky}+T_\text{LNBF}}
	\end{equation}
where the ``hot'' power,
	\begin{equation}
		P_\text{hot}=T_\text{amb}+T_\text{LNBF},
	\end{equation} 
is the power measured when the LNBF is pointed at an ambient source with temperature $T_\text{amb} \sim 300 \text{ K},$ and the ``cold'' power,
	\begin{equation}
		P_\text{cold} = T_\text{sky} + T_\text{LNBF},
	\end{equation} 
is the power measured when the LNBF is pointed at the sky with temperature $T_\text{sky} \sim 3 \text{ K} + \kappa.$ Here, the 3 K is due to the background radiation in the sky, and $\kappa$ is some constant (in units of Kelvin) that characterizes the attenuation of the microwave signal due to atmospheric absorption. 
	\begin{figure}[t]
		\centering
		\includegraphics[width=0.45\textwidth]{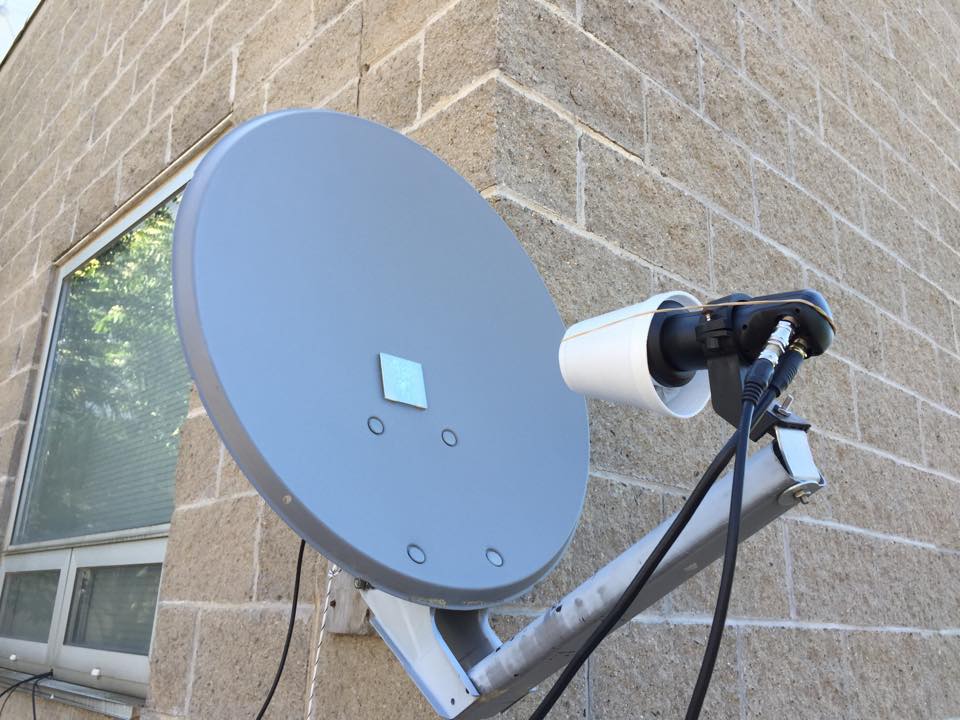}
		\caption{The two--channel ozone spectrometer with an attached microwave absorber during Y-factor testing.}		\label{ytests}

	\end{figure}
	\begin{figure}[h]
		\centering
		\includegraphics[width=0.50\textwidth]{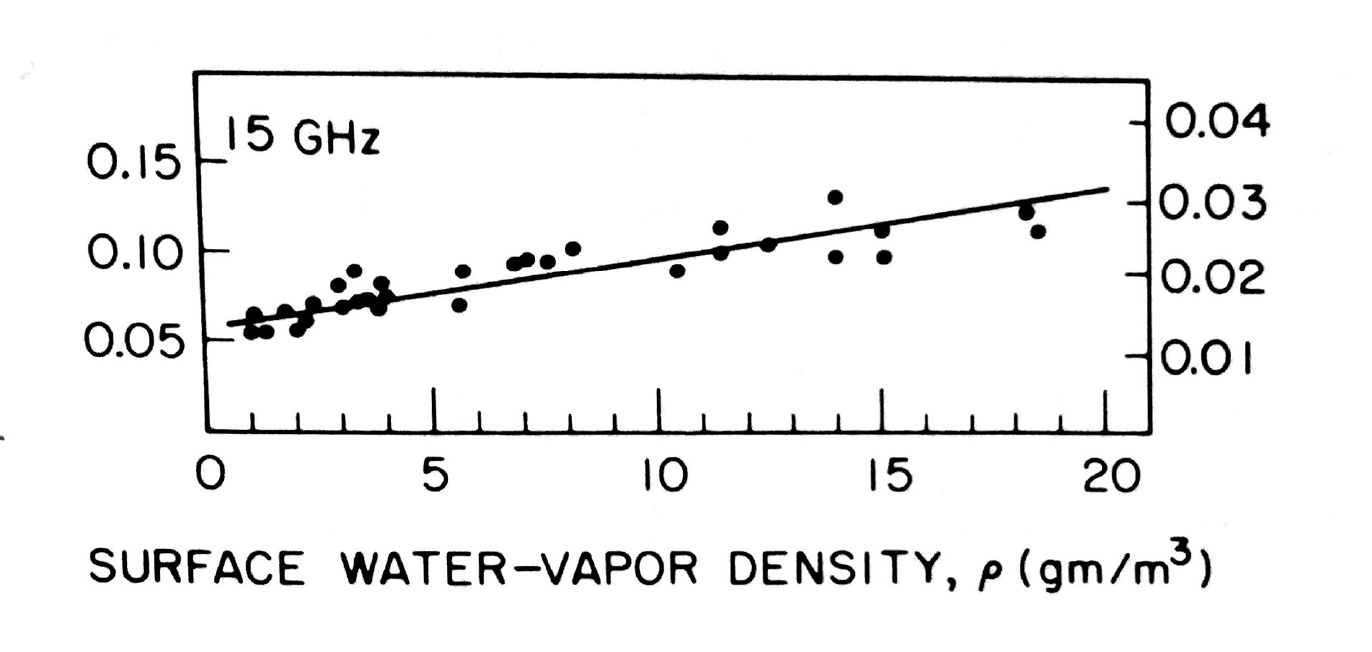}
				\caption{Atmospheric zenith opacity (right-hand vertical axis) and atmospheric decibel attenuation (left-hand) axis versus surface water-vapor density. Note that while Meeks presents data for a 15 GHz signal, the ozone spectrometer is actually tuned to the 11 GHz spectral line.}		\label{fig1}

	\end{figure}	
\begin{figure}[t]
	\centering
	\includegraphics[scale=0.30]{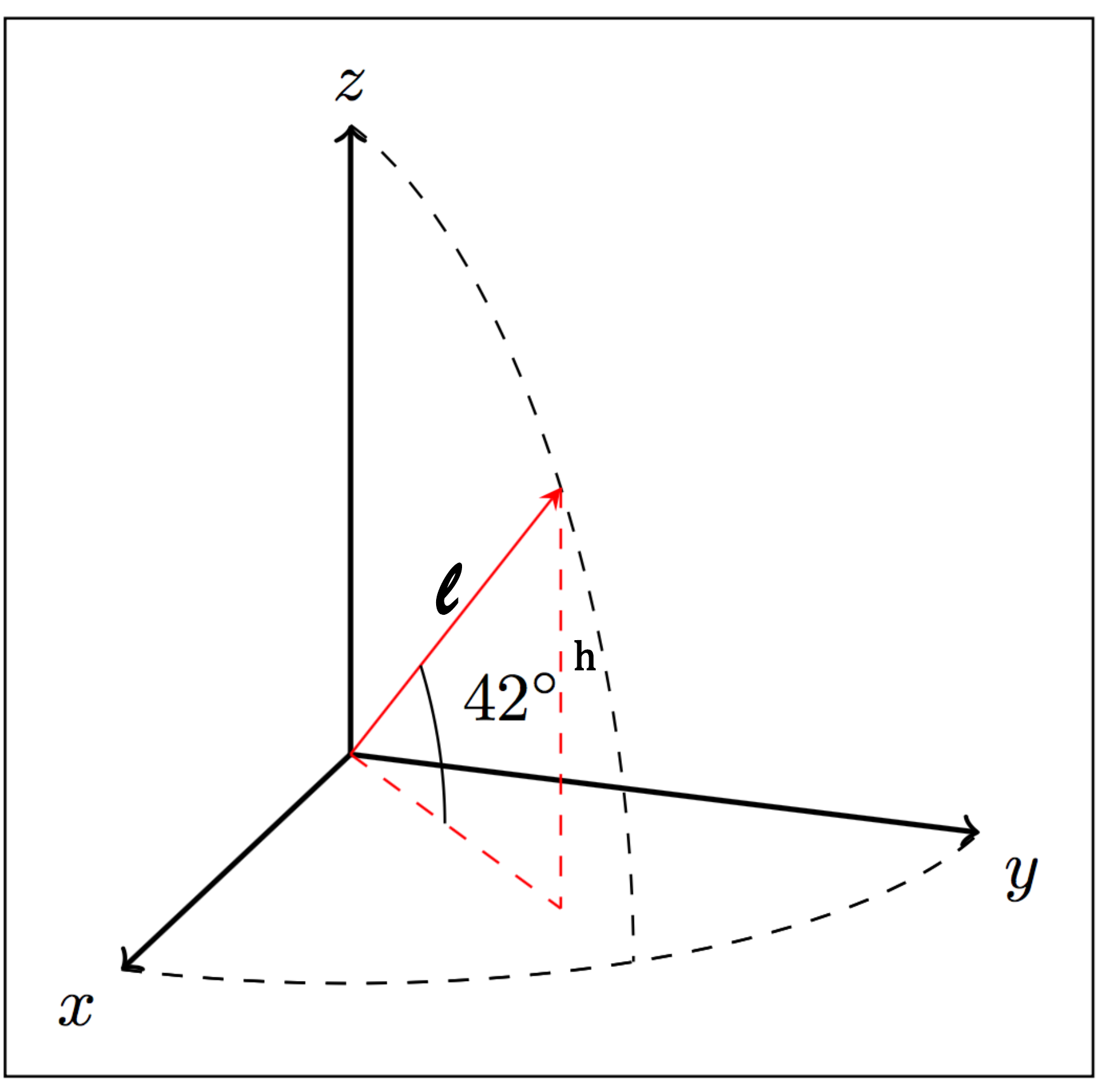}
		\caption{Zenith correction geometry.}\label{Fig8}

\end{figure}
\subsection{Experimental Procedure}
We use a C program to continuously measure and output the power values of the amplifier in dB. Using the results from \textcolor{blue}{Appendix \S\ref{AGEOMAP}}, we point the LNBF at 42 degrees toward the parabolic reflector. The dish itself is then pointed\footnote{In practical runs of the ozone spectrometer, the dish should actually be pointed to around 9 degrees toward the sky. The fact that the dish is pointed 42 degrees toward the sky in this experiment is attributed to human error. \textit{Note}, however, that this does not affect the resulting Y-factor measurements.} 42 degrees toward the sky. This produces a measurement of $P_\text{cold}$ in the program output. Next, a Cuming Microwave pyramid-shaped carbon fiber microwave absorber is wrapped around the feedhorn of the LNBF. Specifically, one pyramid is cut and placed into a styrofoam cup, which can be easily fastened over the LNBF (\textcolor{blue}{Figure} \ref{ytests}). The program then outputs a measurement of $P_\text{hot}$. Since the power units are in log-scale, the Y-factor can be obtained by taking the absolute value of the difference of the two measurements:
	\begin{equation}
		Y\,[\text{dB}]= \text{abs}(\text{dB}_\text{absorber}-\text{dB}_\text{no absorber}).
	\end{equation}
\subsection{Results}

\begin{table}[]
\centering
\begin{tabular}{|l|l|l|l|}
\hline
{\bf Status}                                                                        & {\bf LNB Model}                                                & {\bf Channel 1 (dB)}                                        & {\bf Channel 2 (dB)}                                        \\ \hline
\begin{tabular}[c]{@{}l@{}}Without absorber\\ With absorber\\ Y-factor\end{tabular} & \begin{tabular}[c]{@{}l@{}}Avenger\\ PLL322-S-2\end{tabular}    & \begin{tabular}[c]{@{}l@{}}-46.5\\ -40.1\\ 6.4\end{tabular} & \begin{tabular}[c]{@{}l@{}}-41.0\\ -35.2\\ 5.8\end{tabular} \\ \hline
\begin{tabular}[c]{@{}l@{}}Without absorber\\ With absorber\\ Y-factor\end{tabular} & Chaparral                                                       & \begin{tabular}[c]{@{}l@{}}-40.8\\ -35.6\\ 5.2\end{tabular} & \begin{tabular}[c]{@{}l@{}}-36.5\\ -33.2\\ 3.3\end{tabular} \\ \hline
\begin{tabular}[c]{@{}l@{}}Without absorber\\ With absorber\\ Y-factor\end{tabular} & Octagon 0TL50                                                   & \begin{tabular}[c]{@{}l@{}}-42.5\\ -38.4\\ 4.1\end{tabular} & \begin{tabular}[c]{@{}l@{}}-39.1\\ -34.4\\ 4.7\end{tabular} \\ \hline
\begin{tabular}[c]{@{}l@{}}Without absorber\\ With absorber\\ Y-factor\end{tabular} & \begin{tabular}[c]{@{}l@{}}Star--Com\\ SR-3602 Mini\end{tabular} & \begin{tabular}[c]{@{}l@{}}-39.9\\ -37.1\\ 2.8\end{tabular} & \begin{tabular}[c]{@{}l@{}}-34.2\\ -30.7\\ 3.5\end{tabular} \\ \hline
\begin{tabular}[c]{@{}l@{}}Without absorber\\ With absorber\\ Y-factor\end{tabular} & \begin{tabular}[c]{@{}l@{}}Avenger\\ KSC322-2\end{tabular}      & \begin{tabular}[c]{@{}l@{}}-39.2\\ -34.4\\ 4.8\end{tabular} & \begin{tabular}[c]{@{}l@{}}-37.6\\ -33.1\\ 4.5\end{tabular} \\ \hline
\end{tabular}
\caption{LNBF Y--Factors}
\label{yfacs}
\end{table}
\begin{table}[]
\centering
\begin{tabular}{|l|l|l|}
\hline
{\bf LNB Model}      & {\bf Channel 1 (K)} & {\bf Channel 2 (K)} \\ \hline
Avenger PLL322-S-2    & 37.8                     & 44.4                  \\ \hline
Chaparral             & 52.9                     & 109                  \\ \hline
Octagon 0TL50         & 76.9                     & 62.0                   \\ \hline
Star--Com SR 3602 Mini & 143                    & 99.0                  \\ \hline
Avenger KSC322-2      & 60.6                     & 66.4                   \\ \hline
\end{tabular}
\caption{LNBF Noise Temperatures}
\label{lnbtemp}
\end{table}
\begin{table}[]
\centering
\begin{tabular}{|l|l|l|}
\hline
{\bf LNB Model}      & {\bf Channel 1 (dB)} & {\bf Channel 2 (dB)} \\ \hline
Avenger PLL322-S-2    & 0.53                 & 0.62                 \\ \hline
Chaparral             & 0.73                 & 1.4                \\ \hline
Octagon 0TL50         & 1.0                 & 0.84                 \\ \hline
Star--Com SR 3602 Mini & 1.7                 & 1.3                 \\ \hline
Avenger KSC322-2      & 0.82                 & 0.90               \\ \hline
\end{tabular}
\caption{LNBF Noise Figures}
\label{nfs}
\end{table}
Measurements and computed Y-factors for five different LNB models are given in \textcolor{blue}{Table} \ref{yfacs}. We proceed by calculating the signal attenuation constant $\kappa$. From \cite{EngBox}, we find that the saturation pressure of water vapor is given by the equation
	\begin{equation}
		p_\text{ws} = \frac{\exp\left(77.3450+0.0057T-7235T^{-1}\right)}{T^{8.2}}
	\end{equation}
where $T$ is the dry bulb temperature of moist air. Using a mean temperature of $290.9$ K from weather records  for June 26, 2015, we compute a saturation pressure of $p_{ws} = 2026.79$ Pa \cite{Weath}. The water vapor density (in $\text{kg}\cdot\text{m}^{-3}$) is given by the equation
	\begin{equation}
		\rho_\text{w} = 0.0022\times\frac{p_\text{ws}}{T}
	\end{equation}
and given $T=290.9$ K, we compute a water vapor density of 15.33 $\text{g}\cdot\text{m}^{-3}$ \cite{EngBox}. According to \textcolor{blue}{Figure} \ref{fig1}, a water vapor density of 15.33 $\text{g}\cdot\text{m}^{-3}$ corresponds to an atmospheric zenith opacity $\sim 0.03$. The temperature of the atmosphere follows as
	\begin{equation}
		T_\text{atmosphere} = \kappa = 0.03\times T_\text{amb} \sim 8 \text{ K}.
	\end{equation}
\noindent This yields $T_\text{sky} = 3\text{ K} + 8\text{ K} = 11 \text{ K}.$ However, \textcolor{blue}{Figure} \ref{fig1} cites the \textit{zenith} opacity, and the dish is not pointed at zenith. If the atmosphere has a height $h$ and our dish is pointed at some angle $\theta=42^\circ,$ then the length $\ell$ that forms the hypotenuse of the right triangle in the direction of the dish (see \textcolor{blue}{Figure} \ref{Fig8}), is given by
	\begin{equation}
		\ell = h\csc\theta.
	\end{equation}
We would expect a similar trend in the temperature:
	\begin{equation}
		T'_\text{atmosphere} = T_\text{atmosphere}\,\csc(42^\circ)\sim 12\,\text{K}.
	\end{equation}
Thus, in reality, $T_\text{sky}=3\text{ K}+12\text{ K}=15\text{ K}.$ Now that we have determined $\kappa,$ we return to \textcolor{blue}{Equation} \ref{AOK} to compute the LNB noise temperatures. The results are presented in \textcolor{blue}{Table} \ref{lnbtemp}. Finally, we convert the noise temperature to the noise figure (NF) in decibels, which is defined as
	\begin{equation}
		\text{NF}\,[\text{dB}]\equiv 10\log_{10}\left(\frac{290\text{ K}+T_\text{LNB}}{290\text{ K}}\right).
	\end{equation}
The noise figures for all six LNB models are presented in \textcolor{blue}{Table} \ref{nfs}. 
\subsection{Consensus}
The theoretical noise temperatures for the Star--Com SR-3602 \textit{Mini} are consistent with those measured for the six-channel ozone spectrometer, which was developed by A. E. E. Rogers and E. True \cite{True2014}. This agreement ensures that our calculations are robust. The LNB with the best noise performance is the Avenger PLL322-S-2, which has a noise figure of 0.53 dB on Channel 1 and 0.62 dB on Channel 2. In contrast, the LNB with the worst noise figure appears to be the Star--Com SR 3602 \textit{Mini}, with 1.7 dB on Channel 1 and 1.3 dB on Channel 2. However, upon testing the Avenger PLL322-S-2 in the field, we discovered that there was a significantly higher frequency drift in the Doppler shift of the ozone line at $11$ GHz. This is due to a phase-locked local oscillator inside the Avenger LNB circuitry, which makes it much less stable for long-term data acquisition as our C program is unable to calibrate the LNB frequency at faster rates. In conclusion, we have chosen to continue using the Star--Com SR-3602 \textit{Mini} as our LNB of choice, largely due to its past history of being a stable antenna feed for the ozone spectrometer hardware.
\section{Appendix: Code}
Most of the source files used in this project are publicly available at the ozone spectrometer GitHub repository, which is located at \url{bit.ly/1OsnYf8}. The Python script used to generate the file download from CEDAR Madrigal (\textcolor{blue}{\S\ref{SEC4}}), as well as the file log, is available here: \url{bit.ly/1VL21wS}.	
\section{Acknowledgements}
\textit{O. B. Alam}: I thank A. E. E. Rogers for his invaluable, tireless, and everlasting mentorship over this ten-week Research Experience for Undergraduates (REU) summer program funded by the National Science Foundation (NSF) Grant AST-1156504. All staff at the MIT Haystack Observatory were helpful, especially M. Derome (for eliminating foliage that were in the line of sight of the six--channel spectrometer), P. J. Erickson (for pointing us to the green line data, as well as for program administration), V. L. Fish (for program administration), R. Gaudet (for IT support), L. P. Goncharenko (for a discussion of the physics of atmospheric tides and their relation to the green line data), H. Johnson (for administrative support), F. D. Lind (for Python support), T. Morin (for IT support), K. T. Paul (for administrative support), W. C. Rideout (for CEDAR Madrigal support), J. Vierinen (for Python support), and K. Wilson (for manufacturing the frequency calibration pulse generator circuit). A discussion on 27 July 2015 with S. Kapali, J. Noto, and J. Riccobono, the developers of the Millstone Hill High--Resolution F\'abry--Perot Interferometer, was incredibly helpful. Lastly, I would like to thank Z. J. Hall for elucidating the covariance matrix, Professor C. P. Franck of Cornell University for inspiring me to pursue radio astronomy, and the Massachusetts Institute of Technology and Haystack Observatory Director C. J. Lonsdale for organizing such an incredible summer program.
\bibliographystyle{nature}
\bibliography{OZONE_FINAL}

\begin{thebibliography}{33}
\expandafter\ifx\csname natexlab\endcsname\relax\def\natexlab#1{#1}\fi
\expandafter\ifx\csname url\endcsname\relax
  \def\url#1{\texttt{#1}}\fi
\expandafter\ifx\csname urlprefix\endcsname\relax\def\urlprefix{URL }\fi

\bibitem[{Rogers \emph{et~al.}(2009)Rogers, Lekberg \& Pratap}]{Rogers2009}
Rogers, A. E.~E., Lekberg, M. \& Pratap, P.
\newblock {Seasonal and Diurnal Variations of Ozone Near the Mesopause from
  Observations of the 11.072-GHz Line}.
\newblock \emph{Journal of Atmospheric and Oceanic Technology} \textbf{26},
  2192--2199 (2009).

\bibitem[{Rogers \emph{et~al.}(2012)}]{Rogers2012}
Rogers, A. E.~E. \emph{et~al.}
\newblock {Repeatability of the Seasonal Variations of Ozone near the Mesopause
  from Observations for the 11.072-GHz Line}.
\newblock \emph{Journal of Atmospheric and Oceanic Technology} \textbf{29},
  1492--1504 (2012).

\bibitem[{True(2014)}]{True2014}
True, E.
\newblock Development of a Low Cost Multichannel Spectrometer for the Study of
  Ozone in the Mesosphere.
\newblock Tech. Rep.~76, {MIT Haystack Observatory} (2014).
\newblock \urlprefix\url{bit.ly/1SnJGkS}.

\bibitem[{Rogers \& Alam(2015)}]{VSRT77}
Rogers, A. E.~E. \& Alam, O.~B.
\newblock Two channel spectrometer using Intel NUC.
\newblock Tech. Rep.~77, {MIT Haystack Observatory} (2015).

\bibitem[{{Old Dominion University}(2003)}]{OzoneBook}
{Old Dominion University}.
\newblock Stratospheric Ozone (2003).
\newblock \urlprefix\url{bit.ly/1fowq2S}.

\bibitem[{Gerding \emph{et~al.}(2008)Gerding, H\"offner, Lautenbach, Rauthe \&
  L\"ubken}]{Gerding2008}
Gerding, M., H\"offner, J., Lautenbach, J., Rauthe, M. \& L\"ubken, F.-J.
\newblock Seasonal variation of nocturnal temperatures between 1 and 105 km
  altitude at $54^\circ$ $~\text{N}$ observed by lidar.
\newblock \emph{Atmospheric Chemistry and Physics} \textbf{8}, 7465--7482
  (2008).

\bibitem[{{IUPAC}(2014)}]{IUPAC2014}
{IUPAC}.
\newblock \emph{Compendium of Chemical Terminology}.
\newblock Gold Book (2014).
\newblock \urlprefix\url{bit.ly/1k0yQ5J}.

\bibitem[{Davis \emph{et~al.}(2013)Davis, Du, Smith, Ward \&
  Mitchell}]{Davis2013}
Davis, R.~N., Du, J., Smith, A.~K., Ward, W.~E. \& Mitchell, N.~J.
\newblock {The diurnal and semidiurnal tides over Ascension Island ($8^\circ$
  S, $14^\circ$ W) and their interaction with the stratospheric quasi-biennial
  oscillation: studies with meteor radar, eCMAM and WACCM}.
\newblock \emph{{Atmospheric Chemistry and Physics}} \textbf{13}, 9543--9564
  (2013).

\bibitem[{Sandford \emph{et~al.}(2010)Sandford, Beldon, Hibbins \&
  Mitchell}]{Sandford2010}
Sandford, D.~J., Beldon, C.~L., Hibbins, R.~E. \& Mitchell, N.~J.
\newblock Dynamics of the Antarctic and Arctic mesosphere and lower
  thermosphere - Part 1: Mean winds.
\newblock \emph{Atmospheric Chemistry and Physics} \textbf{10}, 10273--10289
  (2010).

\bibitem[{Rogers(2008{\natexlab{a}})}]{VSRT40}
Rogers, A. E.~E.
\newblock Modeling the diurnal variation of ozone.
\newblock Tech. Rep.~40, {MIT Haystack Observatory} (2008{\natexlab{a}}).

\bibitem[{Fish(2009)}]{VSRT58}
Fish, V.~L.
\newblock Equation for the buildup of ozone at sunset.
\newblock Tech. Rep.~58, {MIT Haystack Observatory} (2009).

\bibitem[{Houghton(2002)}]{PhysAtmos}
Houghton, J.~T.
\newblock \emph{The Physics of Atmospheres} (Cambridge University Press, 2002),
  2nd edn.

\bibitem[{IEEE(2015)}]{mbands}
IEEE.
\newblock Frequency Letter bands (2015).
\newblock \urlprefix\url{bit.ly/1Ml0IAy}.

\bibitem[{{X2 Square Technology}(2014)}]{SR3602}
{X2 Square Technology}.
\newblock {Technical Specifications: Ku-Band $\text{L}$-$40\text{I}$} (2014).
\newblock \urlprefix\url{bit.ly/1RRo8lZ}.

\bibitem[{{MIT Haystack Observatory}(2015)}]{fpi}
{MIT Haystack Observatory}.
\newblock {Atmospheric Optics Facility} (2015).
\newblock \urlprefix\url{bit.ly/1MyzoA3}.

\bibitem[{McLennan \& McLeod(1927)}]{McLennan1927}
McLennan, J.~C. \& McLeod, J.~H.
\newblock On the Wave--Length of the Green Auroral Line in the Oxygen Spectrum.
\newblock \emph{{Proc. Roy. Soc. Lon. A}} \textbf{115}, 515--527 (1927).

\bibitem[{Gao \emph{et~al.}(2012)Gao, Nee \& Xu}]{Gao2012}
Gao, H., Nee, J.-B. \& Xu, J.
\newblock The emission of oxygen green line and density of O atom determined by
  using ISUAL and SABER measurements.
\newblock \emph{Annales Geophysicae} \textbf{30}, 695--701 (2012).

\bibitem[{Phillips \emph{et~al.}(1994)Phillips, Manson, Meek \&
  Llewellyn}]{Phillips1994}
Phillips, A., Manson, A.~H., Meek, C.~E. \& Llewellyn, E.~J.
\newblock A long--term comparison of middle atmosphere winds measured at
  Saskatoon ($52^\circ$ N, $106^\circ$ W) by a medium--frequency radar and a
  Fabry--Perot interferometer.
\newblock \emph{Journal of Geophysical Research} \textbf{99}, 12923--12935
  (1994).

\bibitem[{Steadman \& Thrush(1994)}]{Steadman1993}
Steadman, J.~A. \& Thrush, B.~A.
\newblock A Laboratory Study of the Mechanism of the Oxygen Airglow.
\newblock \emph{Journal of Atmospheric Chemistry} \textbf{18}, 301--317 (1994).

\bibitem[{Hedin \emph{et~al.}(2009)Hedin, Gumbel, Stegman \& Witt}]{Hedin2009}
Hedin, J., Gumbel, J., Stegman, J. \& Witt, G.
\newblock Use of $\text{O}_2$ airglow for calibrating direct atomic oxygen
  measurements from sounding rockets.
\newblock \emph{Atmospheric Measurement Techniques} \textbf{2}, 801--812
  (2009).

\bibitem[{Kohen \emph{et~al.}(2007)}]{Kohen2007}
Kohen, T. \emph{et~al.}
\newblock A Comparison of Long-term Meridional Neutral Winds extracted from
  Arecibo Incoherent Scatter Radar with the Neutral Winds Obtained via Fabry
  Perot Interferometry.
\newblock Radar Conference (IEEE, 2007).

\bibitem[{Xiao \emph{et~al.}(2007)}]{Xiao2006}
Xiao, C.~Y. \emph{et~al.}
\newblock Interpretation of the mesospheric and lower thermospheric mean winds
  observed by MF radar at about $30^\circ$ N with the 2D-SOCRATES model.
\newblock \emph{Advances in Space Research} \textbf{39}, 1267--1277 (2007).

\bibitem[{Brasseur \emph{et~al.}(2000)}]{Brasseur2000}
Brasseur, G.~P. \emph{et~al.}
\newblock Natural and Human-Induced Perturbations in the Middle Atmosphere: A
  Short Tutorial.
\newblock \emph{American Geophysical Union}  (2000).

\bibitem[{Dartt \emph{et~al.}(1983)Dartt, Nastrom \& Belmont}]{Dartt1983}
Dartt, D., Nastrom, G. \& Belmont, A.
\newblock Seasonal and solar cycle wind variations, 80--100 km.
\newblock \emph{Journal of Atmospheric and Terrestrial Physics} \textbf{45},
  707--718 (1983).

\bibitem[{Hedin \emph{et~al.}(1996)}]{Hedin1996}
Hedin, A.~E. \emph{et~al.}
\newblock Empirical wind model for the upper, middle, and lower amtosphere.
\newblock \emph{Journal of Atmospheric and Terrestrial Physics} \textbf{58},
  1421--1447 (1996).

\bibitem[{Drob \emph{et~al.}(2008)}]{Drob2008}
Drob, D.~P. \emph{et~al.}
\newblock An empirical model of the Earth's horizontal wind fields: HWM07.
\newblock \emph{Journal of Geophysical Research} \textbf{113} (2008).

\bibitem[{Yuan \emph{et~al.}(2013)}]{Yuan2013}
Yuan, W. \emph{et~al.}
\newblock FPI observations of nighttime mesospheric and thermospheric winds in
  China and their comparisons with HWM07.
\newblock \emph{Annales Geophysicae} \textbf{31}, 1365--1378 (2013).

\bibitem[{Portnyagin \emph{et~al.}(2004)}]{Portnyagin2004}
Portnyagin, Y. \emph{et~al.}
\newblock Mesosphere/lower thermosphere prevailing wind model.
\newblock \emph{Advances in Space Research} \textbf{34}, 1755--1762 (2004).

\bibitem[{Rogers(2008{\natexlab{b}})}]{VSRT43}
Rogers, A. E.~E.
\newblock Ozone Kinetic Temperature in the Lower Thermosphere.
\newblock Tech. Rep.~43, {MIT Haystack Observatory} (2008{\natexlab{b}}).

\bibitem[{Yuan \emph{et~al.}(2008)}]{Yuan2008}
Yuan, T. \emph{et~al.}
\newblock Climatology of mesopause region temperature, zonal wind, and
  meridional wind over Fort Collins, Colorado ($41^\circ$ N, $105^\circ$ W),
  and comparison with model simulations.
\newblock \emph{Journal of Geophysical Research} \textbf{113} (2008).

\bibitem[{Balanis(1997)}]{Balanis1997}
Balanis, C.~A.
\newblock \emph{Antenna Theory: Analysis and Design}, vol.~2 (John Wiley and
  Sons, Inc., 1997).

\bibitem[{{Engineering Toolbox}(2015)}]{EngBox}
{Engineering Toolbox} (2015).
\newblock \urlprefix\url{bit.ly/1oD60OD}.

\bibitem[{{The Weather Channel}(2015)}]{Weath}
{The Weather Channel} (2015).
\newblock \urlprefix\url{bit.ly/1FKwNu8}.

\end{thebibliography}
\end{document}